\documentclass[final,5p,times,twocolumn,authoryear]{elsarticle}

\usepackage{amssymb}
\usepackage{amsmath}
\usepackage{afterpage}
\usepackage{color}
\usepackage{url}
\usepackage{multicol}

\newcommand{\aapr}{A\&A~Rev.}
\newcommand{\aap}{A\&A}

\newcommand{\apj}{ApJ}

\newcommand{\apjs}{ApJS}
\newcommand{\araa}{ARA\&A}
\newcommand{\mnras}{MNRAS}
\newcommand{\fermilat}{\textit{Fermi}-LAT}

\journal{Astroparticle Physics}

\begin{document}

\begin{frontmatter}



%\title{Assessing capability of machine learning for predicting evolution of enhanced gamma-ray states of active galactic nuclei}
\title{Assessing the capability of random forest to predict the evolution of enhanced gamma-ray states of active galactic nuclei}

\author[afil1]{Tomasz Fidor}
\author[afil1]{Julian Sitarek}

\address[afil1]{Faculty of Physics and Applied Informatics, University of Lodz, ul. Pomorska 149/153, 90-236 \L\'od\'z, Poland}
%\address[afil2]{Department of Astrophysics, The University of Lodz, ul. Pomorska 149/153, 90-236 \L\'od\'z, Poland}
%\affiliation[lab1]{organization={Faculty of Physics and Applied Informatics, University of Lodz},%Department and Organization
%\affiliation[lab]{organization={Department of Astrophysics, University of Lodz},%Department and Organization

\begin{abstract}
  Large fraction of studies of active galactic nuclei objects is based on performing follow-up observations using high-sensitivity instruments of high flux states observed by monitoring instruments (the so-called Target of Opportunity, ToO).
  Due to transient nature of such enhanced states it is essential to quickly evaluate if such a ToO event should be followed.
  We use a machine learning method to assess the possibility to predict the evolution of high flux states in gamma-ray band observed with \fermilat\ in context of following such alerts with current and future Cherenkov telescopes.
  We probe flux and Test Statistic predictions using different training schemes and sample selections. 
  We conclude that a partial prediction of the flux over a time scale of one day with an accuracy of $\sim35\%$ is possible. 
  The method provides accurate predictions of the raising/falling emission trend with $60-75\%$ probability, however deeper investigations shows that this is likely based on typical properties of the source, rather than on the result of most recent measurements. 
\end{abstract}




\end{frontmatter}

%\linenumbers

\section{Introduction} \label{sec:int}
The emission from central regions of some observed galaxies (the so-called Active Galactic Nuclei, AGN) cannot be explained solely by combined  thermal emission  of stars and dust  \citep{ms16,pa17}.
Instead it is expected to originate in non-thermal processes powered by the accretion of matter on the central black hole.
AGN objects with a relativistic jet pointing at the direction of the observer are called blazars \citep{ur91}.
The GeV gamma-ray sky is dominated by AGN objects. 
Over 3100 AGN objects (mainly blazars) has been detected by \fermilat\ in this energy range \citep{ab20}. 
The emission of AGN objects is broadband and has been observed to be highly variable on time scales from minutes to years.
The observed variability of the blazars is often described in terms of noise of a given colour, i.e. different index of power spectral density (see e.g. \citealp{2002MNRAS.332..231U}).
Such an index can show a break reflecting a change of behaviour between long and short time scales, in particular in the gamma-ray band \citep{go18}.
The variability can be also described with an Ornstein-Uhlenbeck model that includes deterministic and non-deterministic part \citep{bu21}. 
The non-linear time series analysis of a sample of bright \fermilat\ blazars shows a combination of deterministic and stochastic processes, with a relevant contribution of the former in some of the source light curves \citep{bh20}.
The recent search of chaos effects in 7-day binned light curves of \fermilat{} blazars shown no evidence of such an effect, suggesting that on such time scales the variability is either stochastic in nature or governed by high-dimensional chaos that can often resemble randomness \citep{os21}. 

Only a handful of sources can be detected in very-high-energy (VHE, $\gtrsim 100$\,GeV) while being in low/typical state \citep[see e.g.][]{an09, acc18, acc20}.
Most of the blazars that  are emitting VHE gamma rays have been detected by following enhanced state observed at lower energies, in particular in GeV range \citep[see e.g.][]{le15,li15,si19}.
The delay in responding to flux enhancements often allows studying only the decaying part of the emission. 
Observations of earlier phases would bring instead additional information not only about cooling of relativistic particles in the sources, but also about their acceleration. 
Identification of high flux state in VHE gamma rays in blazars is important not only to explain the processes happening in those sources, but also because such measurements can be used to probe cosmology and fundamental physics questions, in particular the flux of extragalactic background light \citep[see e.g.][]{acc19} or to search Lorentz invariance violation (LIV) effects \citep[see e.g.][]{ab19}. 
Such types of studies are sensitive on accurate flux measurement at the maximum observed energies from a given event, thus require high photon statistics to reach strong constraints on EBL or LIV. 
Thus, being able to at least partially predict the further evolution of an ongoing flare would allow more efficient target selection for this kind of studies. 

An interesting case showing the potential in predicting gamma-ray flares are gravitationally-lensed blazars, which also provide unique possibilities to constrain the morphology of those sources at angular scales orders of magnitude smaller than the angular resolution of gamma-ray instruments \citep{2018PhR...778....1B}. 
The measured time delay on QSO B0218+357 and a reported  gamma-ray flare of this source was used to predict the occurrence of the second image of the same flare.
Thus, it allowed detection of the source at the VHE gamma ray band, and observe it in a broadband observational campaign already from the onset of the flare \citep{2016A&A...595A..98A}. 
That study had been however marred by the lack of VHE observations of the flare in the first (earlier) image. 
Machine learning flux prediction method might be able to provide faster information about the expected evolution of the source, and hence would aid in catching such flares with VHE gamma-ray in multiple images, opening a possibility for probing their morphology as well as  cosmology \citep{2018PhR...778....1B}.

Thanks to the \fermilat\ monitoring of the sky, the observations of GeV flares are very common. 
In the first 7.4\,years of \fermilat\ data in total 4547 flares have been identified, corresponding to 518 variable gamma-ray sources \citep{ab17}.
Due to limited observation time only a fraction of those alerts can be followed with Imaging Atmospheric Cherenkov Telescopes (IACTs), that are sensitive enough to probe associated VHE gamma-ray emission, but have low duty-cycle and limited field of view.
Moreover, due to aggregation, download and analysis time of \fermilat\ data, as well as visibility constraints, the typical time of follow-up of the flare by IACTs is in between one and two days after the GeV flux measurement that triggered it.
This includes a typical data aggregation time (24\,hrs), transmission and processing of data (6\,hrs, see   \citealp{2018Galax...6..117T}), human evaluation of the measured emission and time needed for the source to become visible by an IACT (typical a few hours). 
Therefore, taking into account the variability of those sources, having a method to extrapolate the flux even over a time scale  of a single day would be appreciated both in the context of currently working IACTs: H.E.S.S. \citep{ah06}, MAGIC \citep{al16a} and VERITAS \citep{weekes2002} and for the future Cherenkov Telescope Array (CTA, \citealp{acha13}).

In the recent years machines learning methods are commonly used in the prediction of time series.  
In particular, the problem of predicting future of astrophysical objects has been tackled e.g. in \cite{spp20}, where authors follow the evolution of galaxy mergers and \cite{ni21} where machine learning is used for prediction of solar flares. 
In the context of \fermilat\ they are typically used for classification of sources (see e.g. \citealp{2016MNRAS.462.3180C,2020arXiv201205251F, 2020MNRAS.493.1926K}).

We investigate how well the GeV variability of AGN objects can be predicted by using a machine learning method: Random Forest Regressor. 
In Section~\ref{sec:data} we describe the used data sample.
The method and pre-processing of the data sample is explained in Section~\ref{sec:meth}.
We present the results of various types of training for different samples in Section~\ref{sec:res}. 
The study is summarised in Section~\ref{sec:conc}.

\section{Data sample} \label{sec:data}
The Large Area Telescope (LAT) is an instrument on board of the \textit{Fermi} satellite.  
LAT is constantly scanning the gamma-ray sky at GeV energies \citep{atw09}.
As a proof-of-concept of the GeV state evolution prediction, we used a publicly available sample of AGN objects monitored\footnote{\url{https://fermi.gsfc.nasa.gov/ssc/data/access/lat/msl_lc/}} by \fermilat{} for which daily automatic analysis is available. 
The total available sample consists of 184 objects, however we excluded non-AGN objects, resulting in 178 sources. 
Moreover, the \fermilat{} monitoring sample has a large number of sources that have shown only episodic emission.
Such sources bring very little information, since they are heavily dominated by non-significant measurements. 
Thus, we further applied selection cuts on the sample to exclude sources that are either too weak or have too low duty-cycle. 
Namely, we take into account only sources which have at least 50 significant flux above 0.1\,GeV measurements (rather than upper limits) and at least 20 pairs of consecutive flux measurements.
The two cuts are motivated by the different training schemes (explained in Section~\ref{sec:rf}), however are strongly correlated.  
The cut values have been selected such that about half of the sources fulfils it. 
Such data selection conserves about a half of the sample (only the more active sources). 
The sources are divided, following the classification in \citet{ab20} into a few classes of objects (see Table~\ref{tab:srcs}): BL Lac objects (BLL), Flat Spectrum Radio Quasars (FSRQ).
The remaining AGN objects (dubbed as ''others'') include blazars of uncertain type (BCU), radio galaxies and Narrow Line Seyfert 1 type galaxies.
Different types of AGN, while sharing some basic features, differ in observed characteristics and underlying physical models (see e.g. the recent reviews \citealp{ms16, pa17} for details).
\begin{table}
  \centering
  \begin{tabular}{c|c|c}
  Type & All & Selected \\ \hline
    BLL & 31 & 22\\
    FSRQ & 118 & 59 \\
%    RG & ... \\
    others & 29 & 7 \\\hline
    all AGN & 178 & 88
  \end{tabular}
  \caption{Number of all available (in automatic daily analysis) and selected sources of different classes in the data sample. 
%In the category of other AGN types we include blazars of uncertain type (BCU), radio galaxies and Narrow Line Seyfert 1 type galaxies..
  }
  \label{tab:srcs}    
  \end{table}
As different sources were added to the \fermilat{} monitoring web page at different times, the duration of the considered light curves varies for different sources (between 338 and 4468 days, with an average of 2972 days for the sources surviving our selection).
% all = agn + bcu + bll + fsrq + rdg 

\section{Prediction method and validation strategy} \label{sec:meth}
We consider two types of characteristic variables in the training: the Test Statistics (TS, \citealp{mat96}) determining the (squared) statistical significance of the signal and  the gamma-ray flux above 0.1\,GeV ($F$).
The uncertainties of the flux are not used in the training, as they do not significantly improve the training (see~\ref{sec:uncert}).

\subsection{Random Forest structure}\label{sec:rf}
We use an implementation of Random Forest (RF) regressor \citep{br01} available in \texttt{scikit-learn} Python package \citep{pe11}.
The RF method has been widely used in astrophysical applications.
In the context of \fermilat{} it has been mainly used for association or classifications of sources (see e.g.  \cite{de14,mi13,saz16}).
While for large training samples and high number of input parameters other methods, such as recurrent neural networks (RNN) ultimately prove more powerful, in the case of a small training samples RF is more likely to reach convergence.  
For example, \citet{2020MNRAS.491.4277M} showed that accuracy of classification of supernova type without galaxy redshift is superior with RNN when using the full sample of $7.1\times 10^{5}$ light curves, but if  $<30\%$ of the sample is used the accuracy with RF method appears to be better than for RNN. 
The limited size of the training data sample of \fermilat\ light curves (88 sources times 2000 days) for this study therefore motivates the usage of the RF method. 
Moreover, the number of features used in the training is also not very large, further supporting usage of a simpler RF method, instead of neural networks that are a commonly used approach for problems with a high number of input variables. 
In addition, the construction of RF allows us to track easily the importance of individual parameters used in the training. 
We have perfomed an additional test using instead a Multi Layer Perceptron Regressor (see~\ref{sec:nn}), however no significant improvement in the prediction power was seen.

We perform two types of training of a RF, the first one is aimed at predicting directly the characteristic variable of the source in the next measurement.
The second is focused on predicting the change of the variable, in particular if the emission will increase/decrease. 

In the first method, dubbed as (i) ''direct'' training,  the data obtained for each source are treated as daily binned time series with $v_i$ being the $i$-th daily measurement of variable $v$.
In order to obtain the prediction $\bar v_i$ for (future) measurement of variable $v_i$ we use $H=20$ earlier measurements, i.e. $v_{i-H}$, $v_{i-H+1}$, ..., $v_{i-1}$.
The complete data sample of $N$ days is divided into two subsamples $1$, ...$M$ and $M+1$, ... $N$.
The first one is used for training the machine learning algorithm and the second one to evaluate its performance.
We select $M\approx 2/3 \times N$, i.e. the training sample is about twice larger then the test sample. 
Note that due to the depth $H$ of the earlier measurements from which a given value is going to be estimated, the actual size of the training samples is $N-H$ and $N-M-H$ respectively. 

In the second method, referred to as (ii) ''differences'' training,  we construct a time series of a change of variable value in consecutive measurements, namely $\Delta v_{i}=v_{i} - v_{i-1}$ (for $i\geq 2$) and apply the training of such a modified time series. 
In this case the training is also applied to predict the parameter value difference to the next measurement. 

Since the TS and flux values in the sample span a range of a few orders of magnitude, to facilitate the training we apply such a transformation\footnote{The transformation in the numerical code are using natural logarithm: $\ln$, however when reporting the results throughout paper, we compute decimal logarithm: $\log$}:
\begin{equation}
v_i^*=
\begin{cases}
\ln (v_i/v_{\min}) & \text{if}\ v_i>v_{\min}  \\
0 & \text{if}\ v_i\leq v_{\min}
\end{cases}\label{eq:trans1}
\end{equation}
where $v_{\min}$ is selected for each type of training variables (TS$_{\min}$=1, $F_{\min}=3\times 10^{-8}\,\mathrm{[cm^{-2}s^{-1}]}$ ). 
For the the case of training on $\Delta v$, which can also be negative we include additional condition. 
\begin{equation}
\Delta v_i^*=
\begin{cases}
\ln (\Delta v_i/v_{\min}) & \text{if}\ \Delta v_i>v_{\min}  \\
-\ln (-\Delta v_i/v_{\min}) & \text{if}\ \Delta v_i<-v_{\min} \\
0 & \text{if}\ v_{\min}>\Delta v_i>-v_{\min},
\end{cases}\label{eq:trans2}
\end{equation}
The selected value of $F_{\min}$ is of the order of the lowest daily fluxes that might be measured with \fermilat{}, the typical measured fluxes are about an order of magnitude higher.  
The transformations are reversible for $v_i>v_{\min}$ and $|\Delta v_i|>v_{\min}$. 
If the flux measurement is reported as an upper limit, the upper limit value is used in the training as a variable as it would be a flux measurement\footnote{We have tested different approaches of dealing with upper limits in the time series, e.g. setting them instead to $v_{\min}$, however this resulted in a strong bias of the resulting training. See~\ref{sec:opt} for details.}. 

We use 100 estimators (RF trees). 
%In order to avoid overtraining over a limited data sample we set the final node size to be smaller than 6 
In order to avoid overtraining over a limited data sample we set the final node size to be smaller than 6 
(further splitting down to single events has been also tested, but did not result in large changes of the performance of the method). 
No other constrain on the RF tree (i.e. a maximum depth of the trees or maximum number of final nodes) is applied. 
At each RF splitting step the best variable (by maximising the decrease of the mean squared error) is chosen from the list of randomly selected square root of the total number of variables. 

\subsection{Training and testing strategies}

At the testing stage, depending on the training method, we obtain either (i) directly the (logarithm of) the value of the characteristic variable, or (ii) the (logarithm of) difference to the last measurement. 
To obtain the expected change in the first case, or the actual value in the second one we use the measured value from the last available measurement. 

We test also different training schemes (i.e. sample selection/division): (A) independent training of each source, and (B) combining of all the samples. 
Finally, we also consider combining samples for all sources belonging to a given class (see Table~\ref{tab:srcs}). 

\subsection{Performance measures}
To evaluate the prediction power and performance of the method we use a few quantities -- performance measures.
We define Mean absolute error of logarithm as $\mathrm{MAE}=\langle |\ln \bar v_i-\ln v_i|\rangle$. 
%Mean relative error is defined as $\mathrm{MRE}=\langle |(\bar v_i-v_i)/v_i|\rangle$. 
If the true value is only reported as a flux upper limit it is not possible to compare it with the prediction, therefore such a point is not included in the evaluation of MAE for flux. 
Similarly, if the most recent point available for the prediction is an upper limit, the ''differences'' training would not be able to estimate properly the upcoming flux. 
Therefore to allow a fair comparison between both training methods we exclude such points from MAE calculations for both training schemes.
While the presence of upper limits does not affect directly the training for the TS variable, we clip the values at TS$>1$ in the training accuracy evaluation. 

We also compute a fraction $f_{u/d}$ of measurements in which the method properly predicts increase or decrease of the value. 
We also consider separately $f_u$ and $f_d$, the fraction of properly predicted raises and drops of the flux respectively. 
In the case of the ''direct'' training, the increase is defined if the predicted value is larger than the last measurement. 
For the ''differences'' training the predicted raise or fall of the emission is the sign of the output of prediction. 
In the case of flux prediction, because of multiple upper limits  in the data sample, we take a similar approach as in the calculation of MAE parameters. 
If the two consecutive time bins give a flux measurement we use the difference to derive if the flux has increased or decreased. 
%If the first measurement was an upper limit and the second resulted in flux measurement, we consider it a raise of the flux, and the other way around a flux measurement followed by an upper limit is considered a flux decrease. 
But if any (or both) of the two consecutive time bins provide only a flux upper limit, such a point is not taken into account in the calculation of accuracy of the up/down prediction.  

\section{Results}\label{sec:res}
We test the performance of the prediction method on different data samples. 

\subsection{Single source}
We first apply the proposed method to a single source.
We have selected 3C454.3, a bright and strongly variable FSRQ. 
We perform both types of training: on direct values and on differences.

\begin{figure}[t]
  \includegraphics[width=0.48\textwidth]{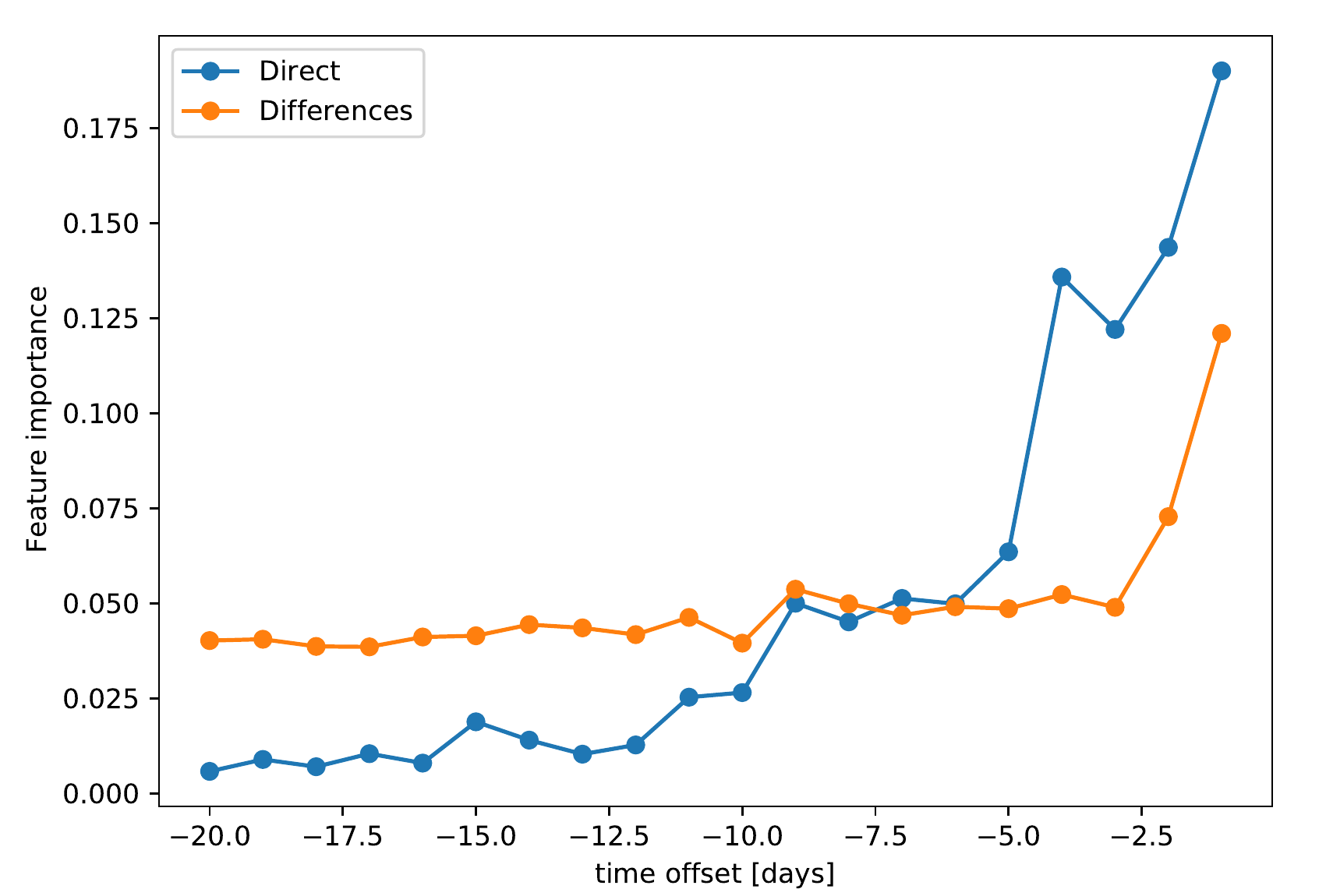}
  \caption{Relative importance of individual parameters used in the training (i.e. representing the values of the variable measured a given number of days earlier).
  Training is done for TS values of 3C454.3 data. 
  }\label{fig1}
\end{figure}

In Fig.~\ref{fig1} we show the importance of individual parameters (i.e. $H$ last measurements of the variable) used in the training.
It is measured as the normalized total reduction of the mean square error during the training. 
As expected, the most important parameters in the training are the most recent measures of quantity to be predicted. 
Interestingly, for the direct training the values spanning about a week before prediction are strongly used. 
On the other hand if the training is done on value differences mainly the last two flux differences are used. 

\begin{figure*}[t!]
\centering
\includegraphics[width=0.48\textwidth]{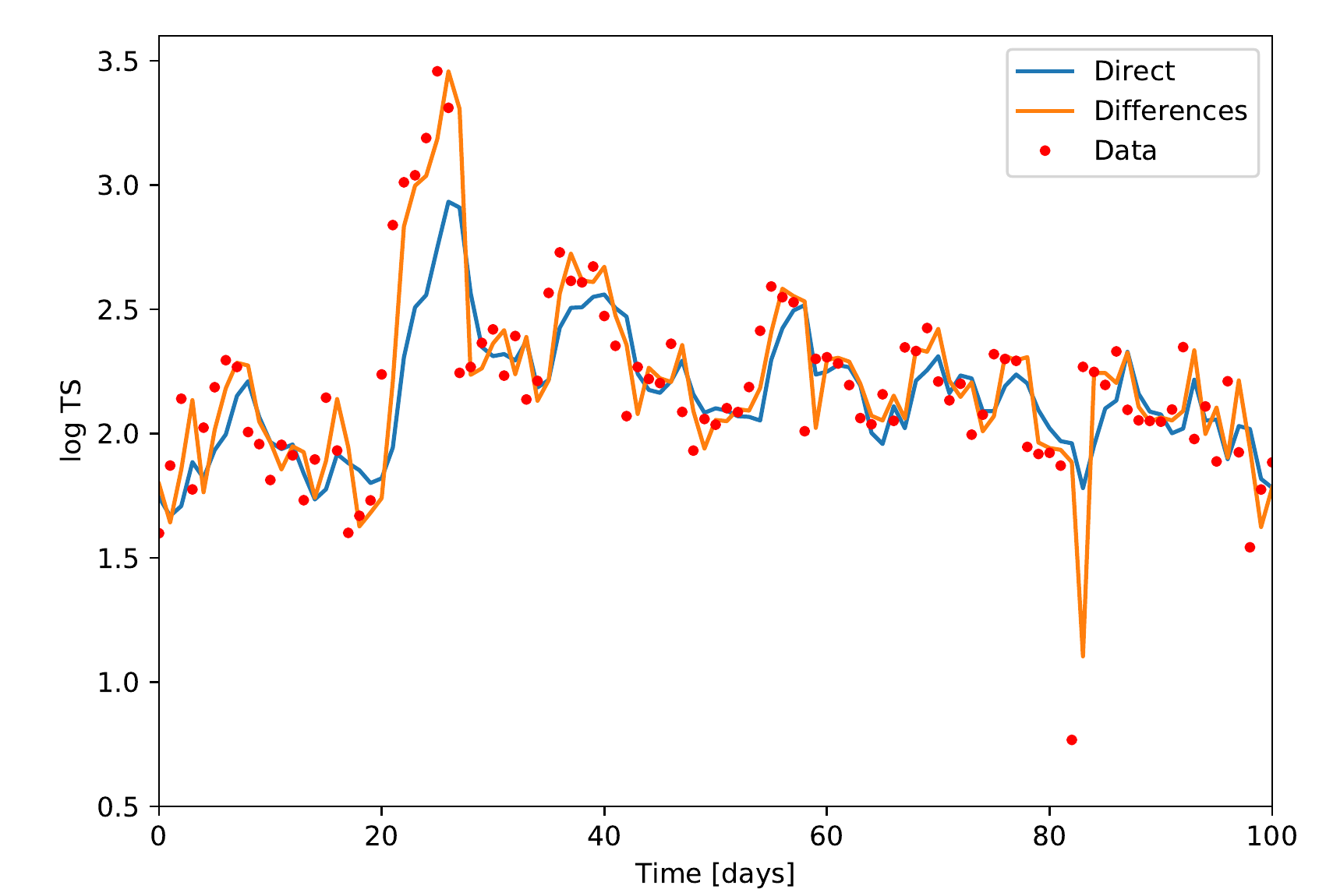}
\includegraphics[width=0.48\textwidth]{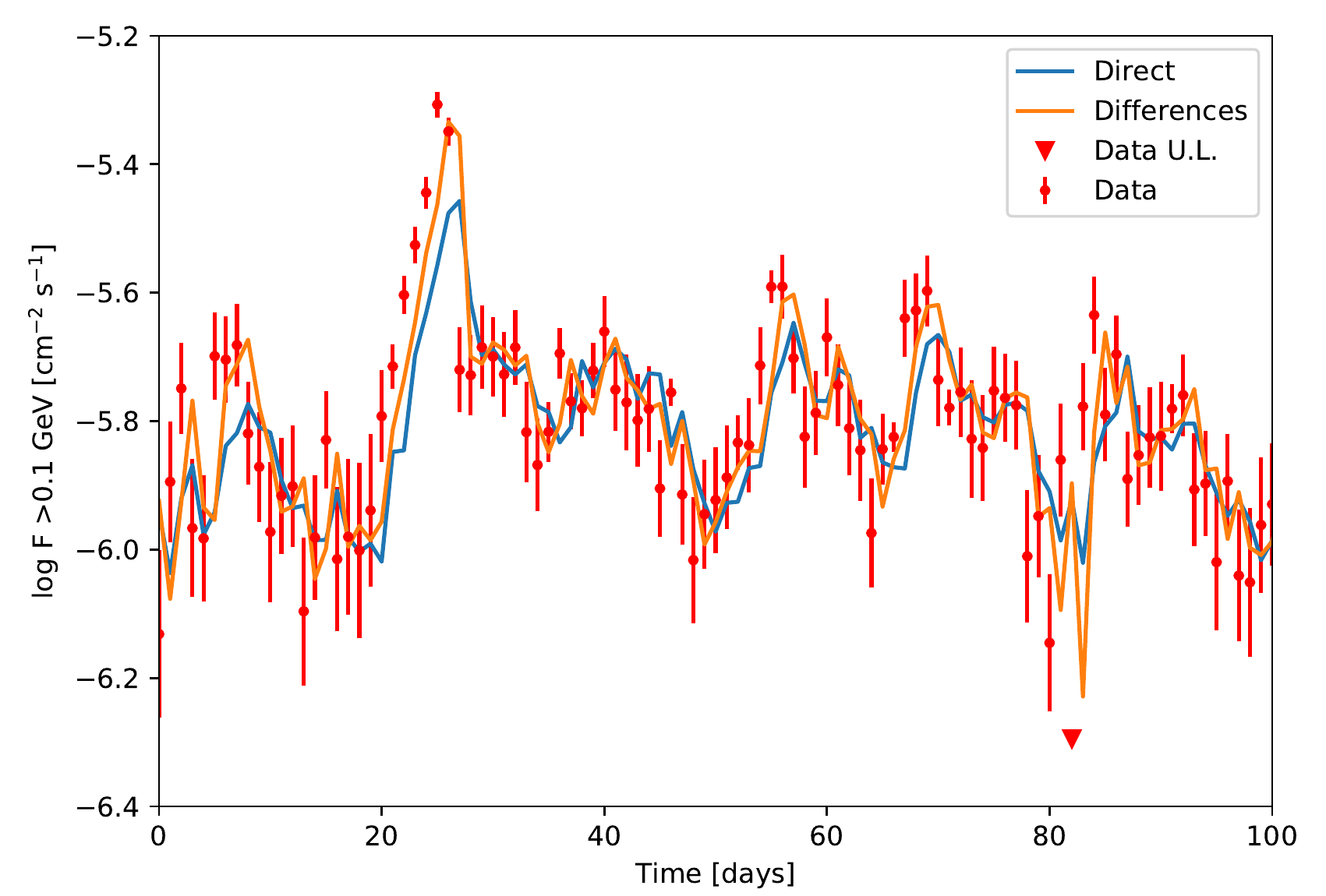}
\includegraphics[width=0.48\textwidth]{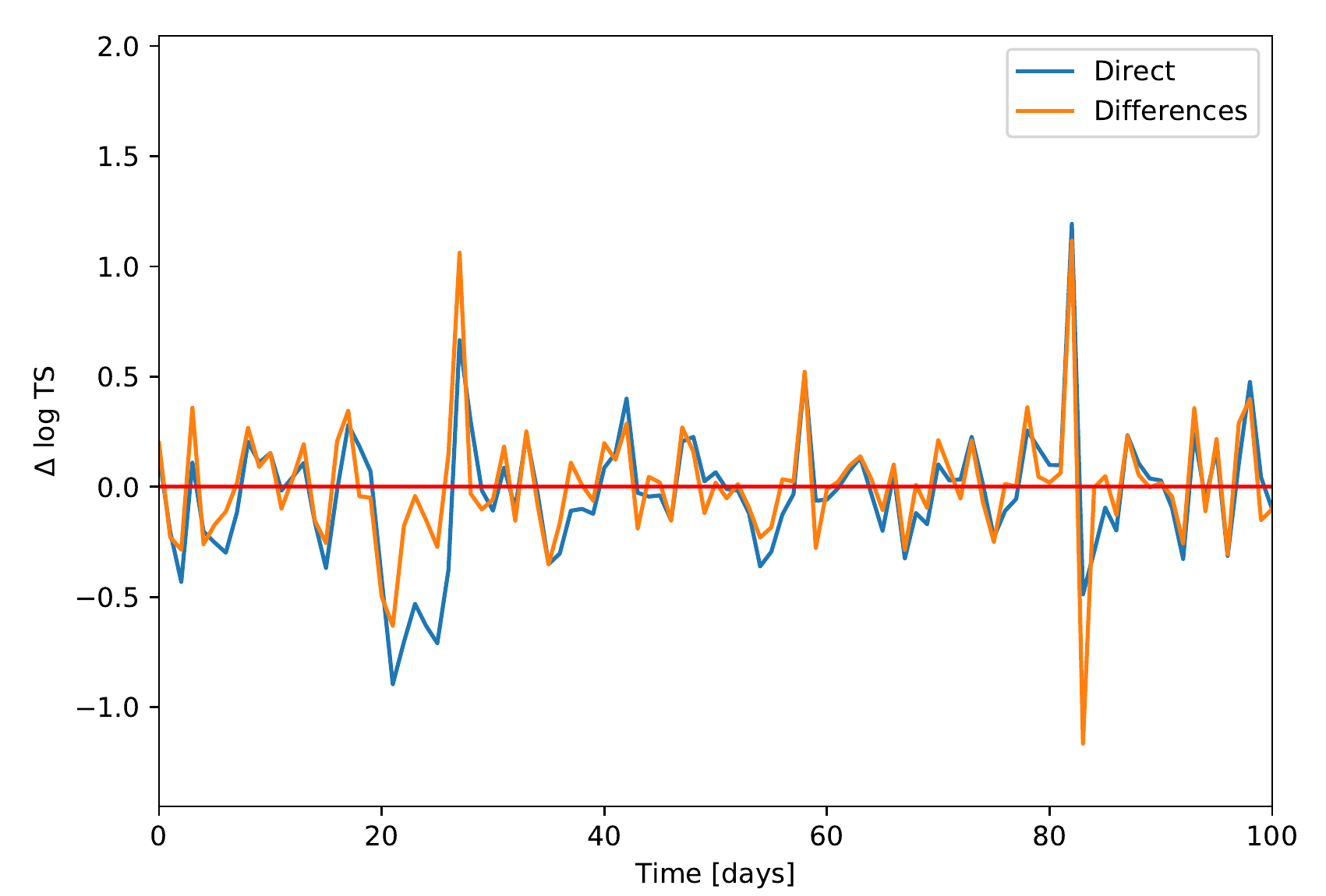}
\includegraphics[width=0.48\textwidth]{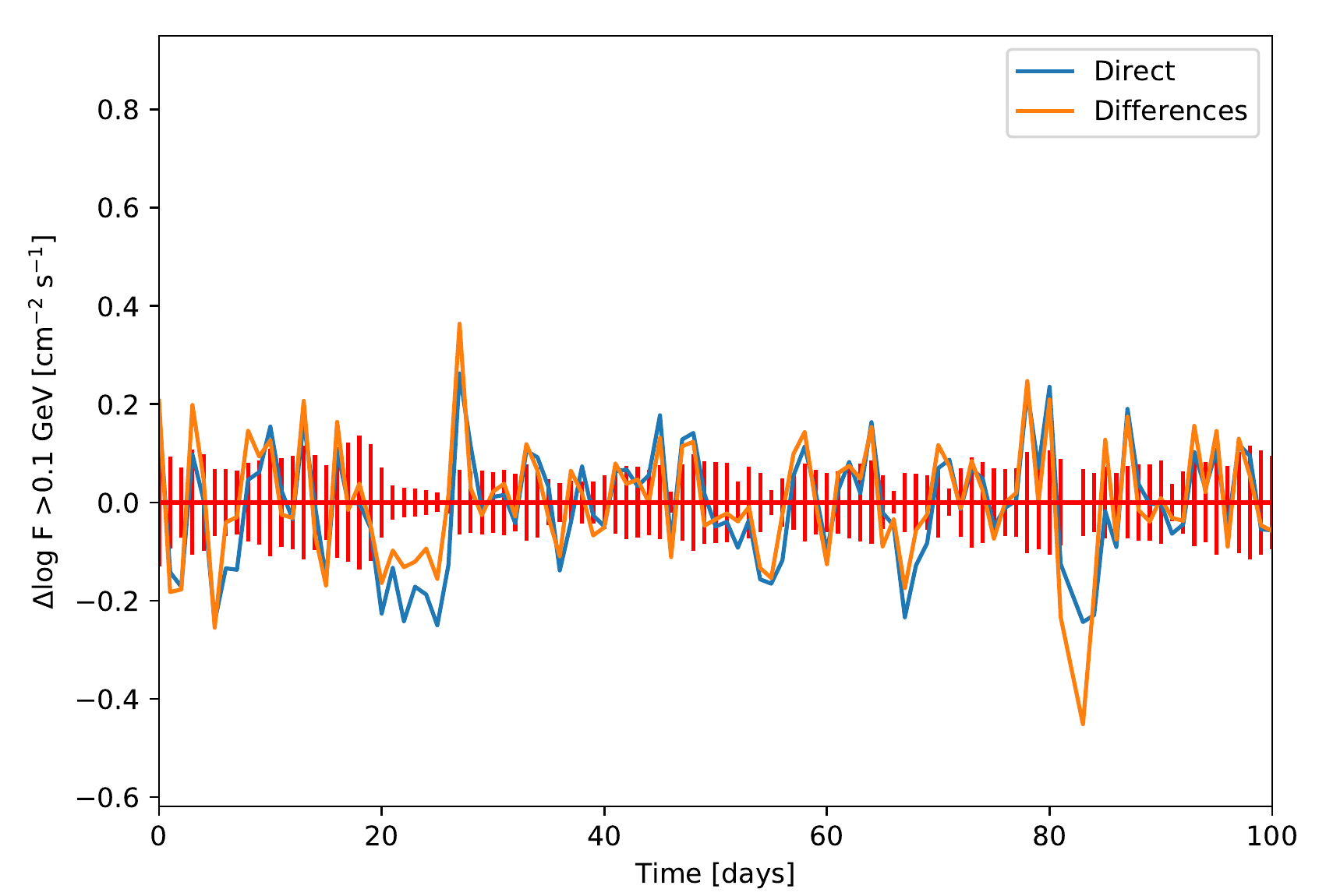}
  \caption{Comparison of the prediction (using direct training, in blue, and training on differences, in orange) with the \fermilat{} measurements (red markers) for 3C454.3.
  Left panels shows TS, right panels flux. 
  The bottom panels show the difference between the prediction and the measured value. 
  For clarity only first 100 points of the test sample are shown. 
}\label{fig2}
\end{figure*}

\begin{figure*}[t!]
\includegraphics[width=0.48\textwidth]{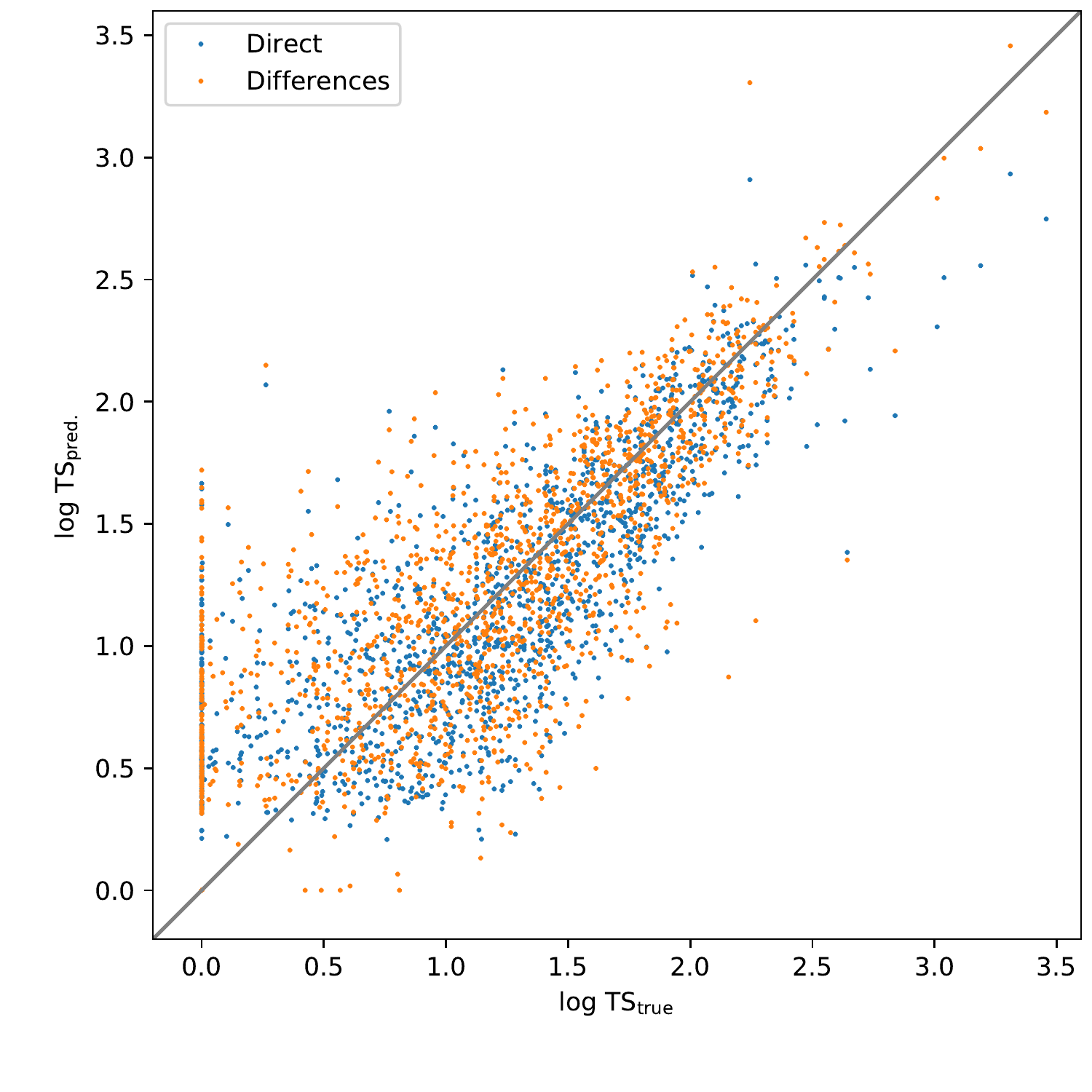}
\includegraphics[width=0.48\textwidth]{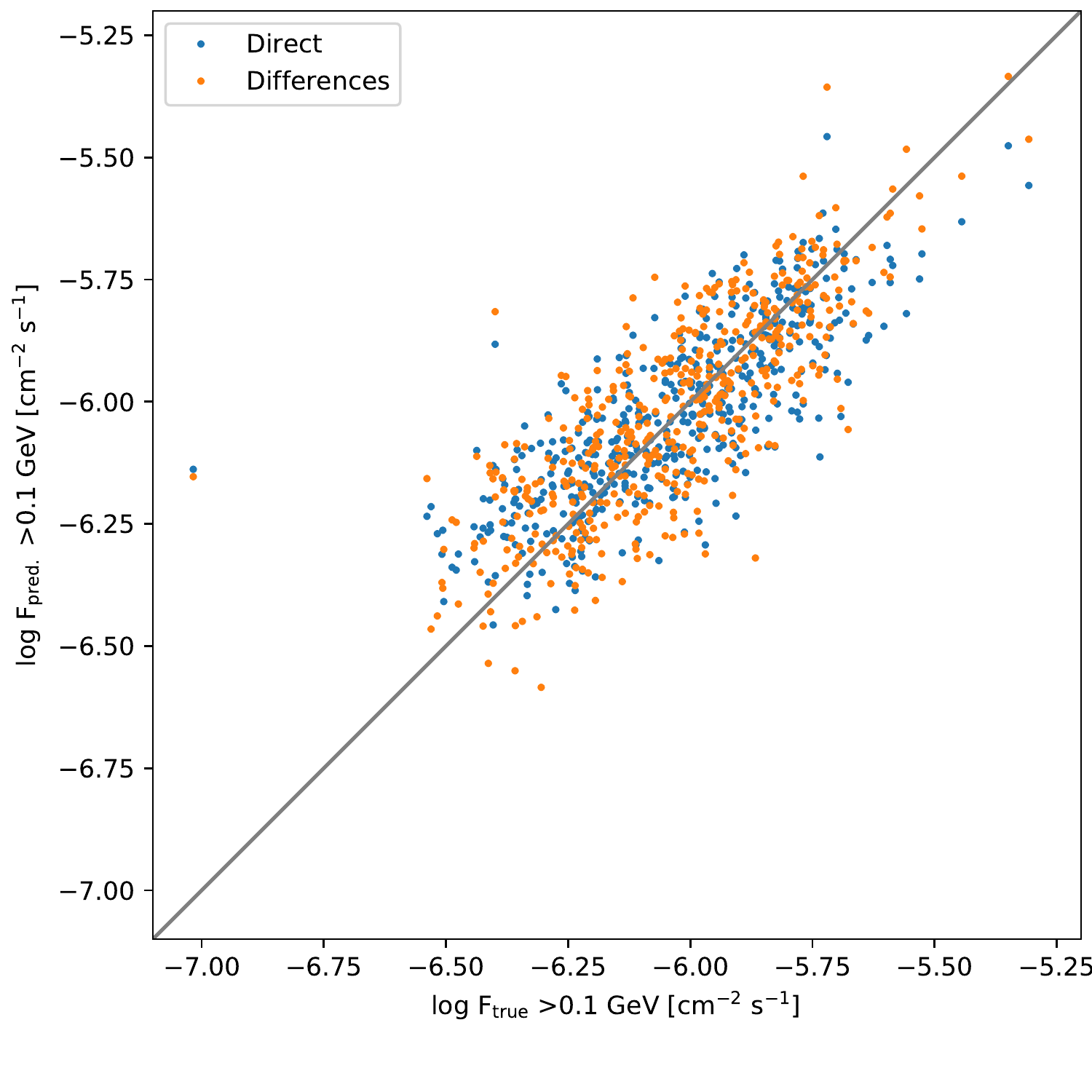}
  \caption{Predicted vs measured value of TS (left panel) and flux (right panel) 
  for direct training (blue dots) and training on differences (orange dots) for 3c454.3.
  In the left panel the points for which true TS is lower than $TS_{\rm min}=1$ are shown as vertical feature at $\log TS_{\rm true}=0$. 
  In the right plot the points for which either the last available measurement or the measurement that is being predicted resulted in an upper limit are omitted.}\label{fig3}
\end{figure*}

In Fig.~\ref{fig2} we compare the measured and predicted light curves and in Fig.~\ref{fig3} we confront the predictions of the algorithm with the consecutive measurement for both considered variables.
Both methods are able to roughly reproduce the evolution of the light curve.
During the periods of slow evolution of the light curve the difference of the measured and predicted value are comparable to the measurement uncertainty, however during rapid variability periods the differences are larger. 
The predicted vs measured distribution is concentrated in a band along the perfect prediction line. 
%However in the case of direct training there is some bias visible, in particular for weaker fluxes,  underestimating the prediction. 

As expected the predictions are limited by the stochastic component of the variability. 
The predictions in general are more accurate for long, smooth evolution than for sharp, one day-time scale features. 
In particular sharp raises result in underestimation of the value. 
Similarly, since the prediction is mainly driven by most recent measurements, sharp drops of the emission show a kind of "hysteresis" effect resulting in overestimation of the value. 
Also in the case of flux prediction if the last available measurement is a strong upper limit, then this results in lowering of the prediction in the next point. 

\subsection{Full sample}
\begin{figure*}[t]
\includegraphics[width=0.48\textwidth]{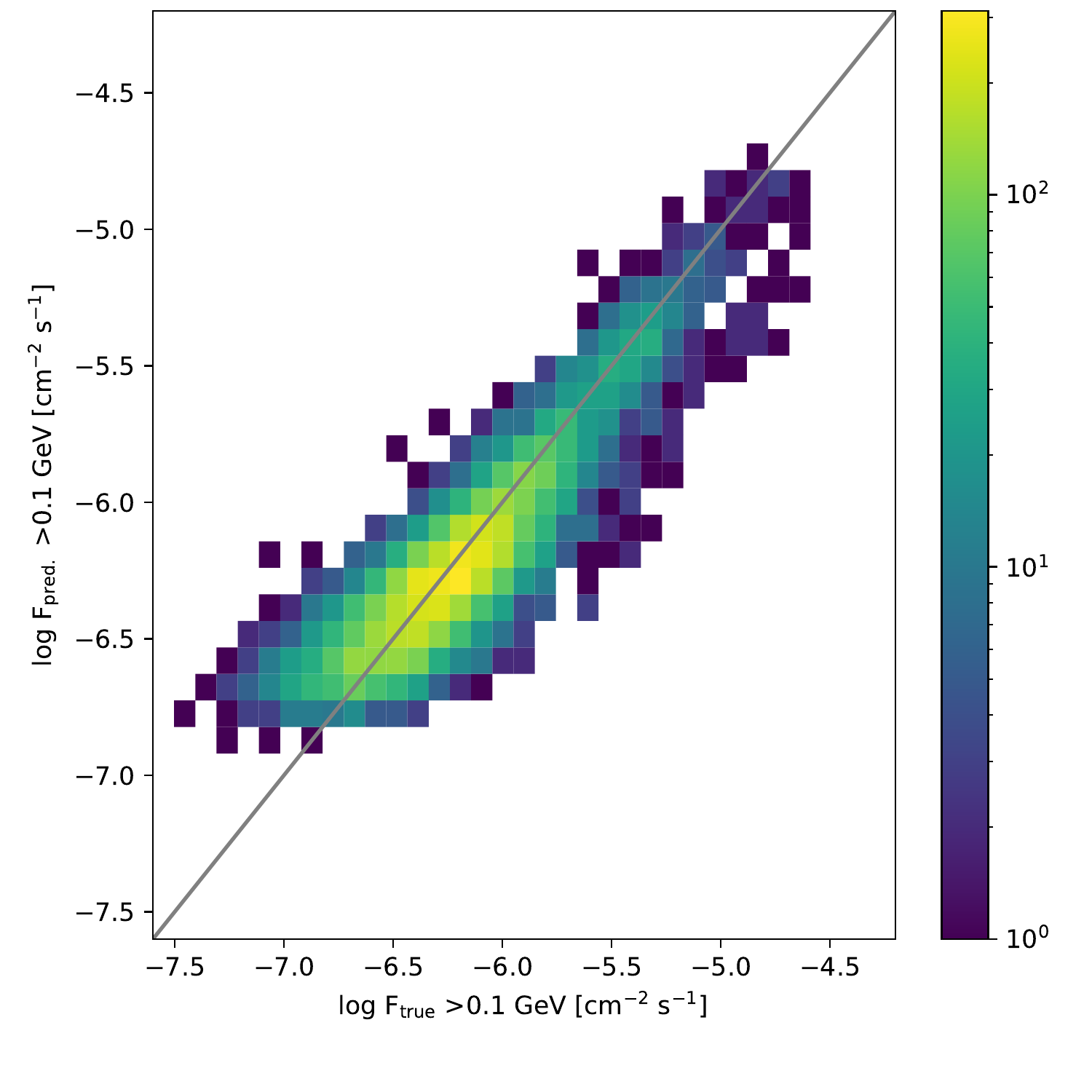}
\includegraphics[width=0.48\textwidth]{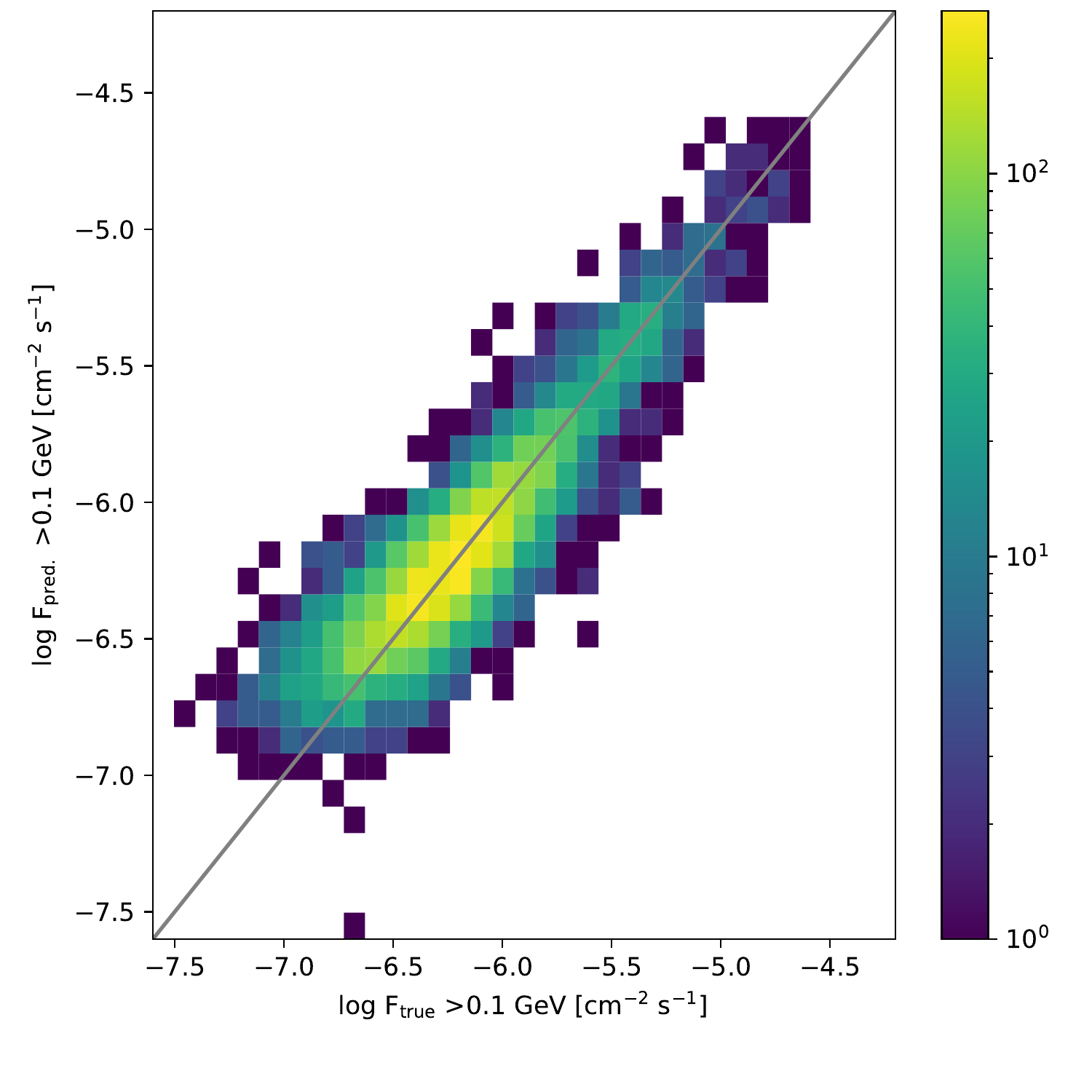}
  \caption{
  Distribution of predicted vs measured value of flux for direct training (left panel) and training on differences (right panel) for the whole sample. 
  The points for which either the last available measurement or the measurement that is being predicted resulted in an upper limit are omitted.}\label{fig4}
\end{figure*}

Next we apply the method to the full sample of 88 sources. 
We either perform a dedicated training for each of the sources (A), or, combining statistics, a single training for all the sources (B). 
In the latter case we first prepare for each source $H$-tuples of previous measurements, used as training parameters, and then use the collection of such bunches from all the sources in single RF. 
This way we avoid a situation in the accuracy evaluation in which the prediction would be performed over a part of a light curve from one source and part from the next source. 
In Fig.~\ref{fig4} we present the dependence of the predicted vs measured value of flux for the whole sample.
The distribution shows a similar behaviour to the one presented for a single source in Fig.~\ref{fig3}, namely the distribution is mainly populated in a broad line around the perfect prediction. %, with some bias (mainly for ''direct'' training) seen at low fluxes. 
At the lowest measured flux values an overestimation bias is visible. 
It can be caused by the fact that upper limit values are used together with flux measurements in the training. 
For the sources close to the detection threshold, such upper limits values will be close, however on average larger than the true flux, thus resulting in overestimation of the prediction. 
We investigate the effect of the upper limits in the training sample in \ref{sec:noul}. 
%It can be caused by the fact that for flux measurements close to the detection, the upper limits used in the training will overestimate the flux prediction. 

\begin{figure*}[t]
\includegraphics[width=0.48\textwidth]{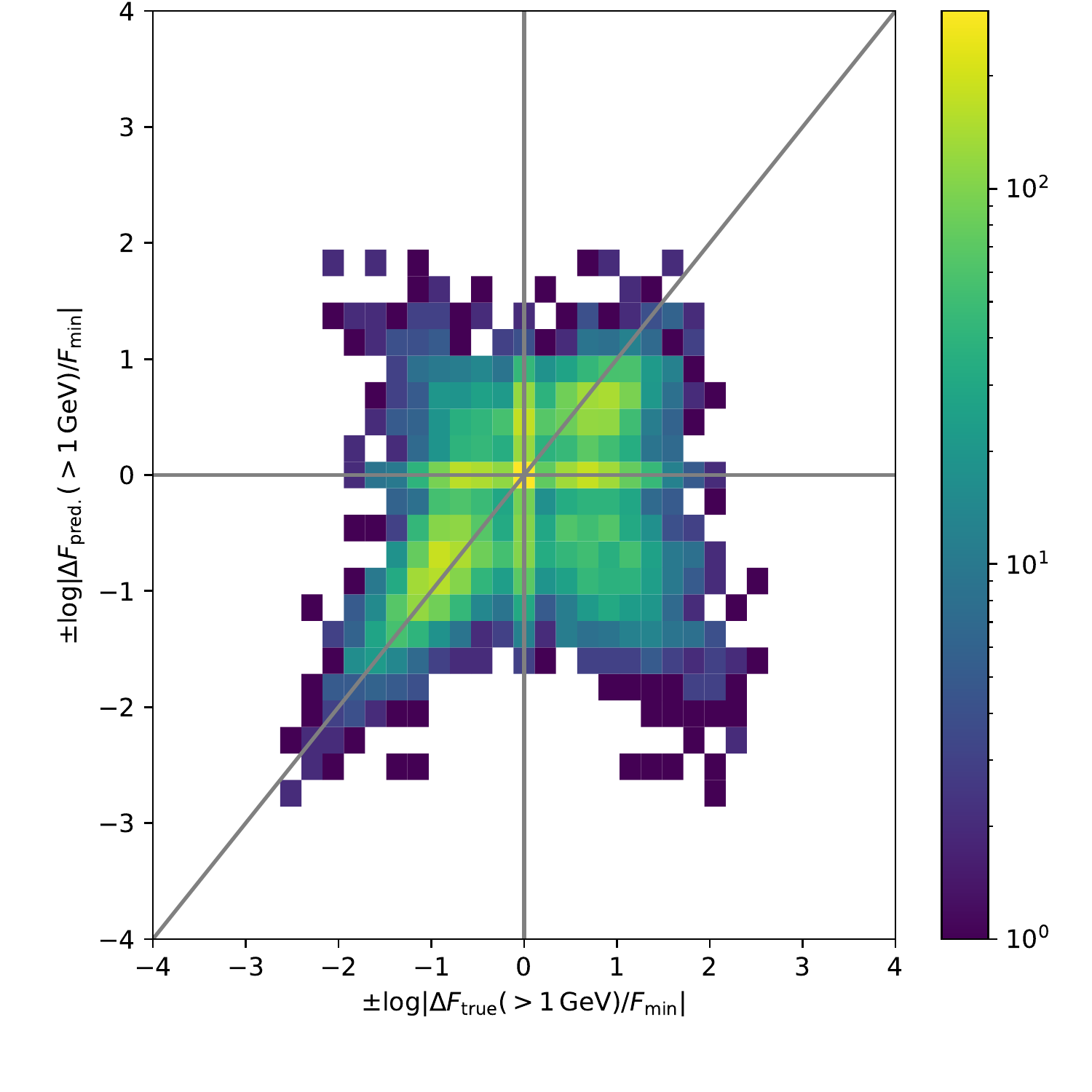}
\includegraphics[width=0.48\textwidth]{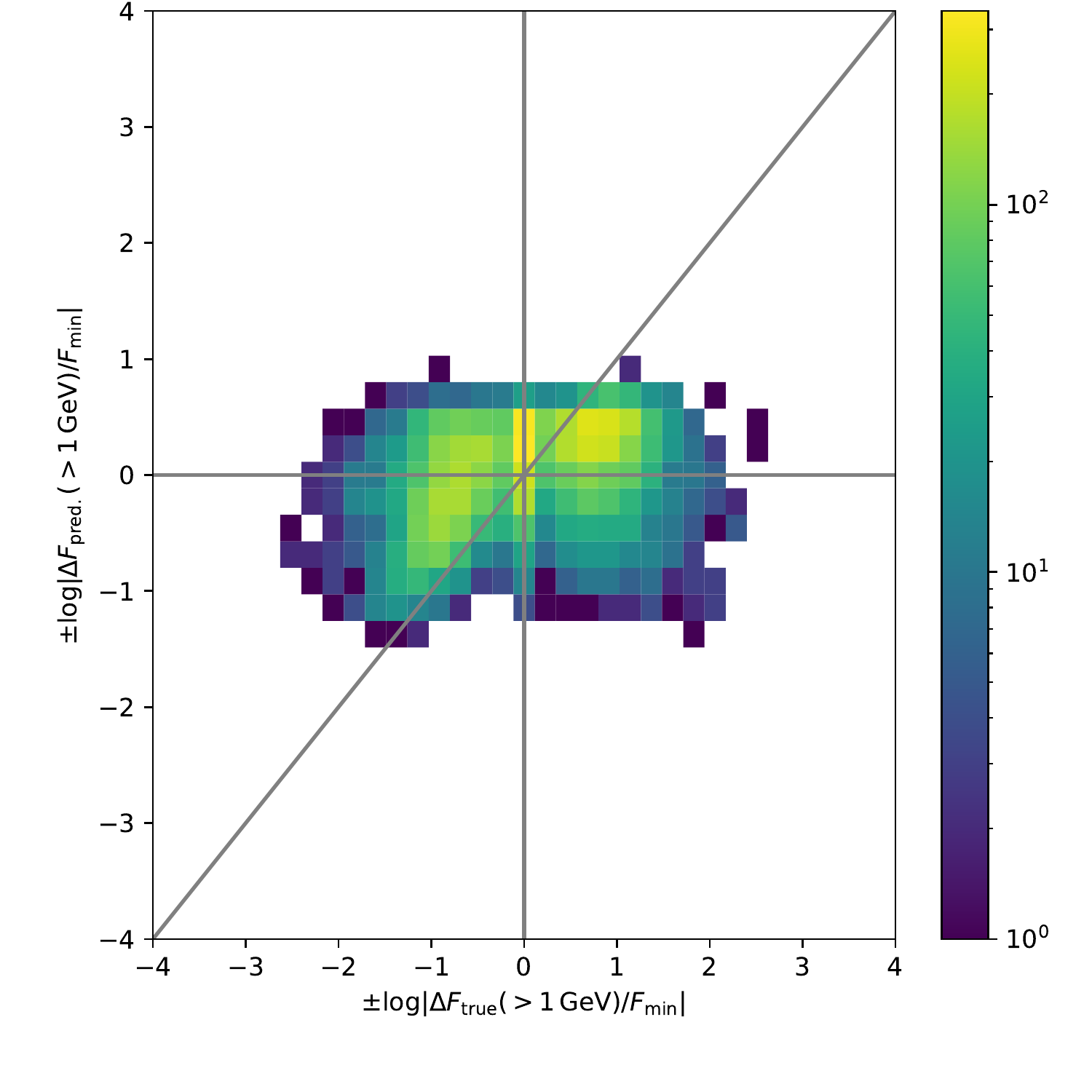}
  \caption{
  Difference to the previous measurement of the predicted value plotted as a function of such a difference of measured value for training on flux of all sources.
  Left and right panels show ''direct'' and ''differences'' training respectively. 
  The points for which either the last available measurement or the measurement that is being predicted resulted in an upper limit are omitted.}\label{fig4b}
\end{figure*}

In Fig.~\ref{fig4b} we confront the predicted raise/fall of the emission with the actually observed one for both types of training. 
For visibility the flux difference is plotted after the transformation of Eq.~\ref{eq:trans2}. 
The correlation on predicted flux differences is much less pronounced than the correlation on the actual values shown in Fig.~\ref{fig4}.
This is understandable as the uncertainty of the prediction often will be larger than the flux increase/decrease with respect to previous measurement. 
Nevertheless, the distributions show an excess of entries in top-right and bottom-left quarters of the plot (i.e. properly predicted upward/downward trend) with respect to the two remaining quarters. 
The analogical distributions as those shown in Fig.~\ref{fig4} and Fig.~\ref{fig4b}, but for TS instead of flux show similar general features, therefore are not presented. 

We summarise the performance measures for all the applied training schemes in Table~\ref{tab:pars}.
\begin{table*}
  \centering
  \begin{tabular}{l|l|r|r|r|r | r|r|r|r }
\multicolumn{2}{c}{} & \multicolumn{4}{|c|}{TS} & \multicolumn{4}{|c}{$F$}  \\ \hline
    Train method & Train sample & MAE   & $f_{u/d}$[\%] & $f_u$[\%] & $f_d$[\%] & MAE & $f_{u/d}$[\%] & $f_u$[\%] & $f_d$[\%] \\\hline
%    (i) Direct   &  (A) single  & 0.798 & $73.6\pm0.2$ & $65.8\pm0.3$ & $81.5\pm0.2$ & 1.093 &  $75.9\pm0.3$ & $53.6\pm0.6$ & $98.3\pm0.1$ \\
%    (i) Direct   &  (B) all  & 0.768 & $74.0\pm0.2$ & $65.6\pm0.3$ & $82.3\pm0.2$& 0.787 & $75.8\pm0.3$ &  $53.8\pm0.6$ & $98.0\pm0.2$ \\\hline
%    (ii) Differences   &  (A) single & 0.878 & $69.4\pm0.2$ & $79.1\pm0.2$ & $59.7\pm0.3$ & 0.373 & $70.9\pm0.4$ & $67.3\pm0.5$ & $74.6\pm0.5$\\
%    (ii) Differences   &  (B) all  & 0.865 & $71.8\pm0.2$ & $79.6\pm0.2$ & $64.0\pm0.3$ & 0.372 &  $74.2\pm0.3$ & $66.7\pm0.5$ &  $81.6\pm0.4$\\
    (i) Direct   &  (A) single  & 0.798 & $73.6\pm0.2$ & $65.8\pm0.3$ & $81.5\pm0.2$ &  0.372 & $64.6\pm0.5$ & $55.2\pm0.8$ & $74.0\pm0.7$ \\
    (i) Direct   &  (B) all  & 0.768 & $74.0\pm0.2$ & $65.6\pm0.3$ & $82.3\pm0.2$&  0.308 & $64.9\pm0.5$ & $56.5\pm0.8$ & $73.4\pm0.7$ \\\hline
    (ii) Differences   &  (A) single & 0.878 & $69.4\pm0.2$ & $79.1\pm0.2$ & $59.7\pm0.3$ & 0.306 & $62.3\pm0.5$ & $74.4\pm0.7$ & $50.1\pm0.8$ \\
    (ii) Differences   &  (B) all  & 0.865 & $71.8\pm0.2$ & $79.6\pm0.2$ & $64.0\pm0.3$ & 0.299 & $64.1\pm0.5$ & $72.6\pm0.7$ & $55.5\pm0.8$ \\
  \end{tabular}
  \caption{Comparison of the performance parameters obtained for different methods of training. }
  \label{tab:pars}    
  \end{table*}
For all combinations of training method and sample selection (i.e. single sources trained separately or all sources combined) the efficiency of the trend prediction (measured by $f_{u/d}$ parameter) is about 60--75\%. 
Since the values are above 50\% it shows that a partial prediction of the emission trend is possible. 
Those results are also in line with excess in the top-right and bottom-left quarters of distributions in Fig.~\ref{fig4b}. 
The trend prediction is slightly more accurate for TS ($\sim 70-75\%$) than for $F$ ($\sim65\%$). 
While the values $f_{u/d}$ are roughly comparable for the ''direct'' and ''differences'' training, the values of $f_u$ and $f_d$ show a strong dependence on the training type. 
For both TS and $F$ parameters the ''direct'' training is more efficient in correctly predicting drops, while the ''differences'' training achieves a better fraction of properly predicted raises of emission. 
Since the raising and falling parts of the light curve do not need to be symmetric, it is reasonable to expect differences between those two parameters. 
%Also the earlier discussed biases can affect them, in particular the occurrence of upper limits if the flux prediction. 
%In particular the underestimation of the fluxes in the ''direct'' training in some cases (visible e.g. in top panel of Fig.~\ref{fig4}) biases as well the expected trend direction resulting in $f_d$ reaching $\sim98\%$. 
The training method itself might also introduce a bias, favouring raise or fall by the two methods. 
Nevertheless, the fractions of properly predicted raises and drops of the flux, $f_u$ is still above 50\% for all the combinations, showing that the prediction is still useful for predicting both types of behaviour. 
%Interestingly, while for ''direct'' training the above-mentioned bias results in $f_d>f_u$ for both TS and flux training, for training on differences the situation is more complicated.
%While for TS  emission raises are more often correctly predicted than drops, for flux it is the other way around. 

The achieved accuracy of the value prediction, measured with MAE, is similar for both methods. 
In the case of TS, the value of MAE of $\sim0.8-0.9$  corresponds to an accuracy of about a factor of two. 
In the case of flux prediction, the achieved accuracy is much better, reaching 0.3 (corresponding to an average accuracy of about 35\%). 
The better accuracy for the flux prediction than for TS is likely connected with the fact that the $F$ prediction is limited only to significant flux measurements.
In contrast, the RF for TS variable is trying to predict also weak hints of emission (with TS$<25$), which are strongly affected by measurement fluctuations. 
%For the total used sample, and the used assumption of excluding from the test sample points in which either the predicted or the last available value are upper limits, 

The difference between performing one universal training combining all the samples into one single training, or performing separate training for each source separately is not large. 
Nevertheless, the former results in an improvement of nearly all investigated performance measures. 
It is mostly pronounced in MAE value for direct training of flux ($\Delta$MAE=0.064), but the same effect is also present in the other method/variable combinations ($\Delta$MAE=0.007--0.030). 
The improvement of the accuracy with a combined training is likely caused by rather small samples for individual sources, too small for efficient training. 
We investigate further methods to improve the values of the performance measures and possible biases in \ref{sec:opt}.

\subsection{Source types}
\begin{figure*}[t!]
\includegraphics[width=0.33\textwidth]{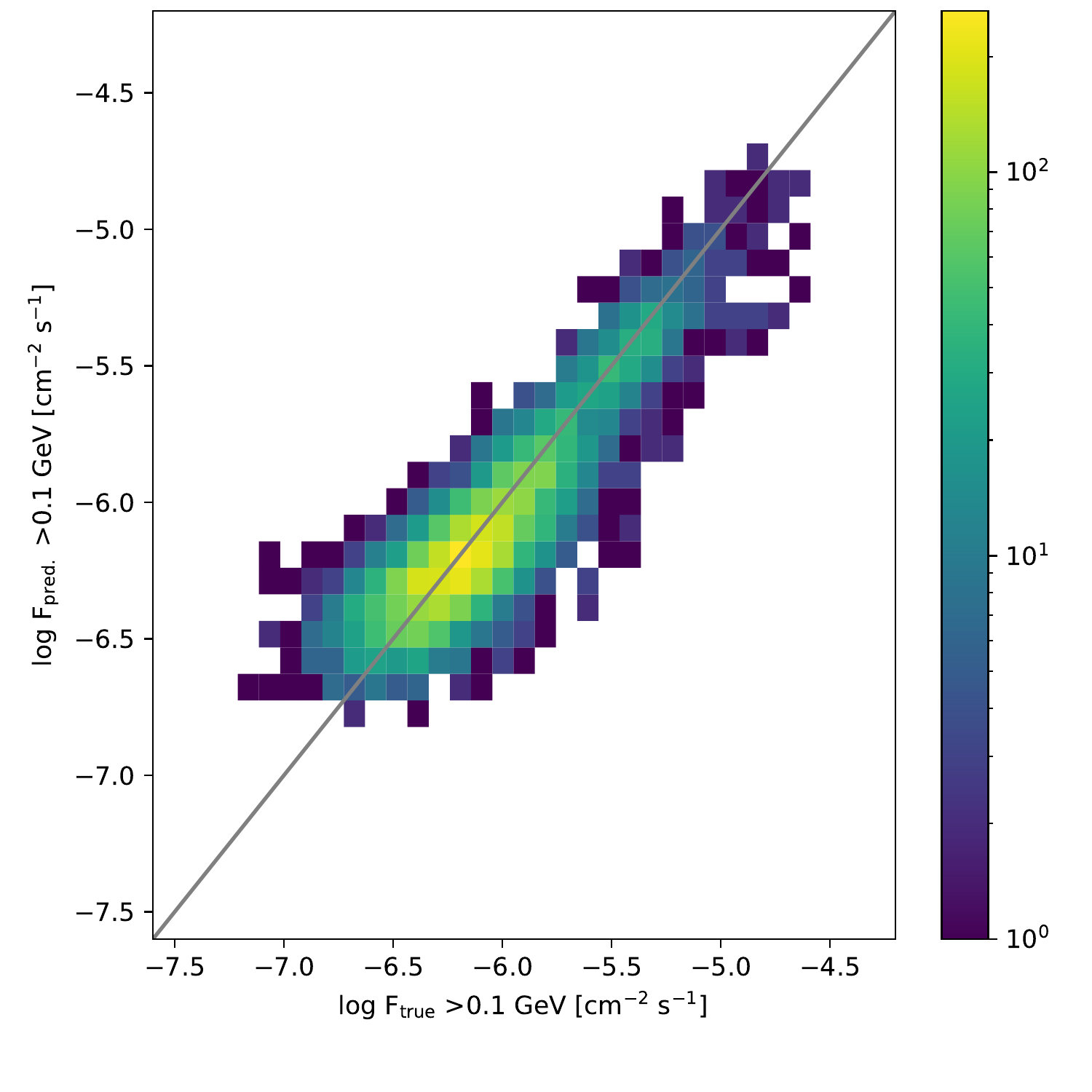}
\includegraphics[width=0.33\textwidth]{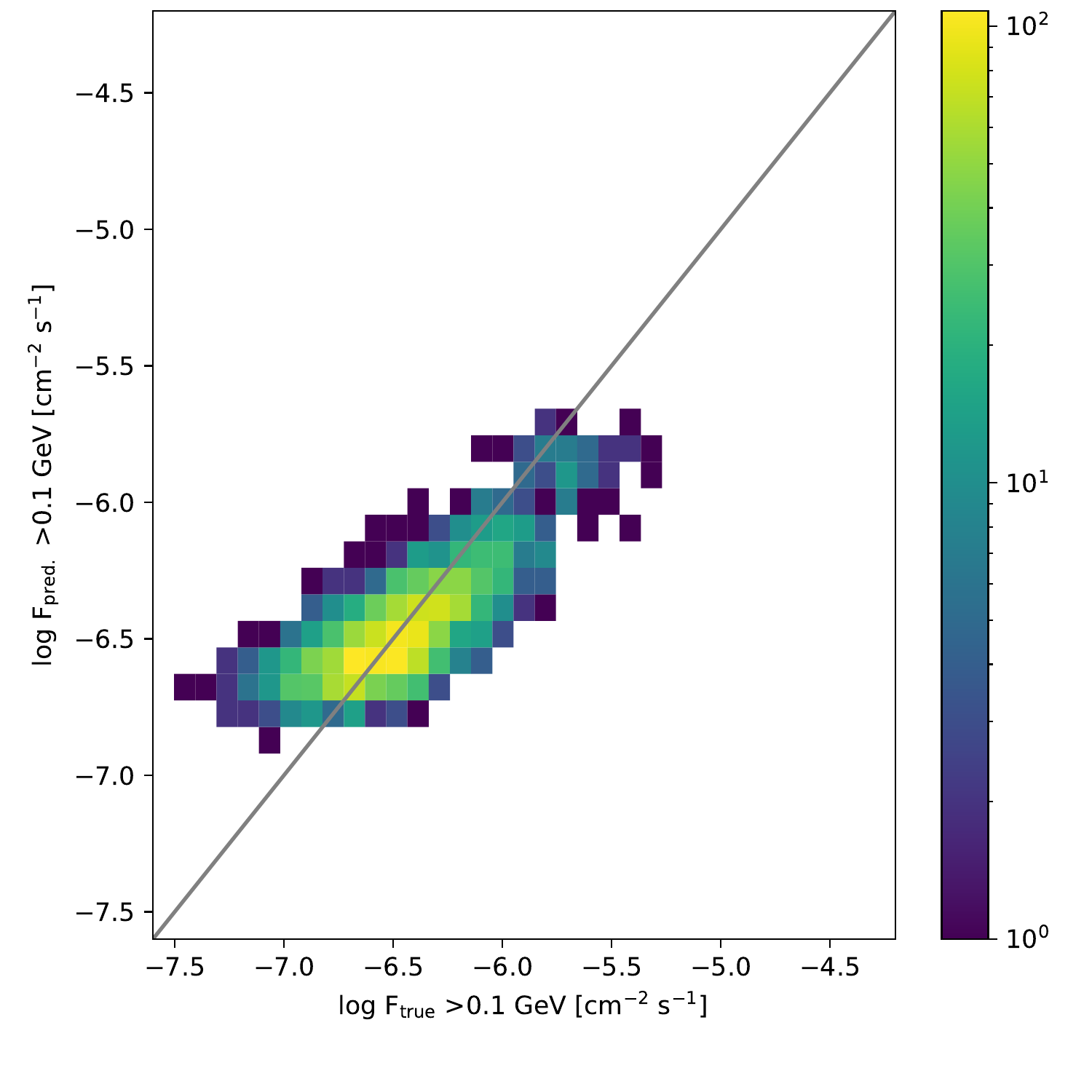}
\includegraphics[width=0.33\textwidth]{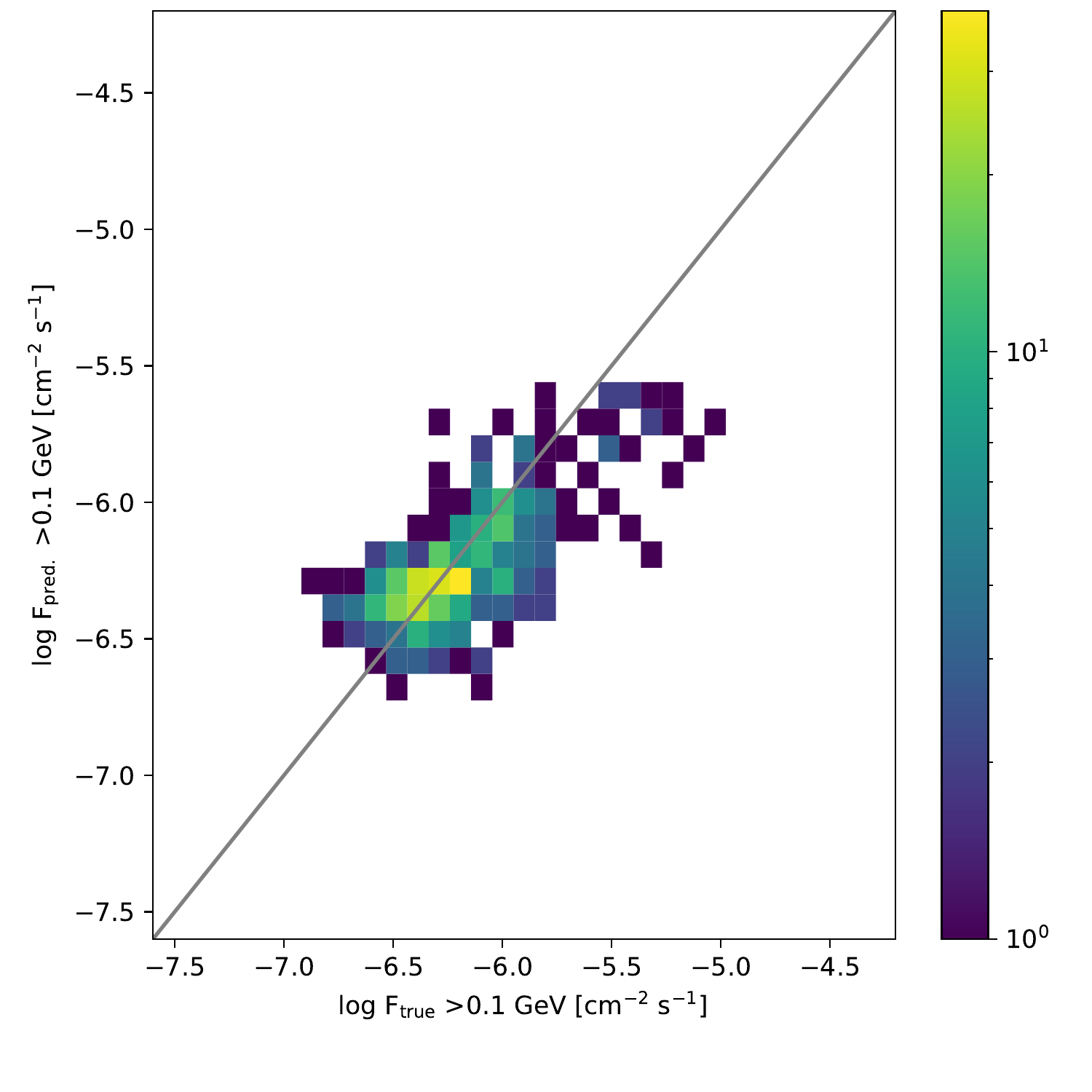} \\
\includegraphics[width=0.33\textwidth]{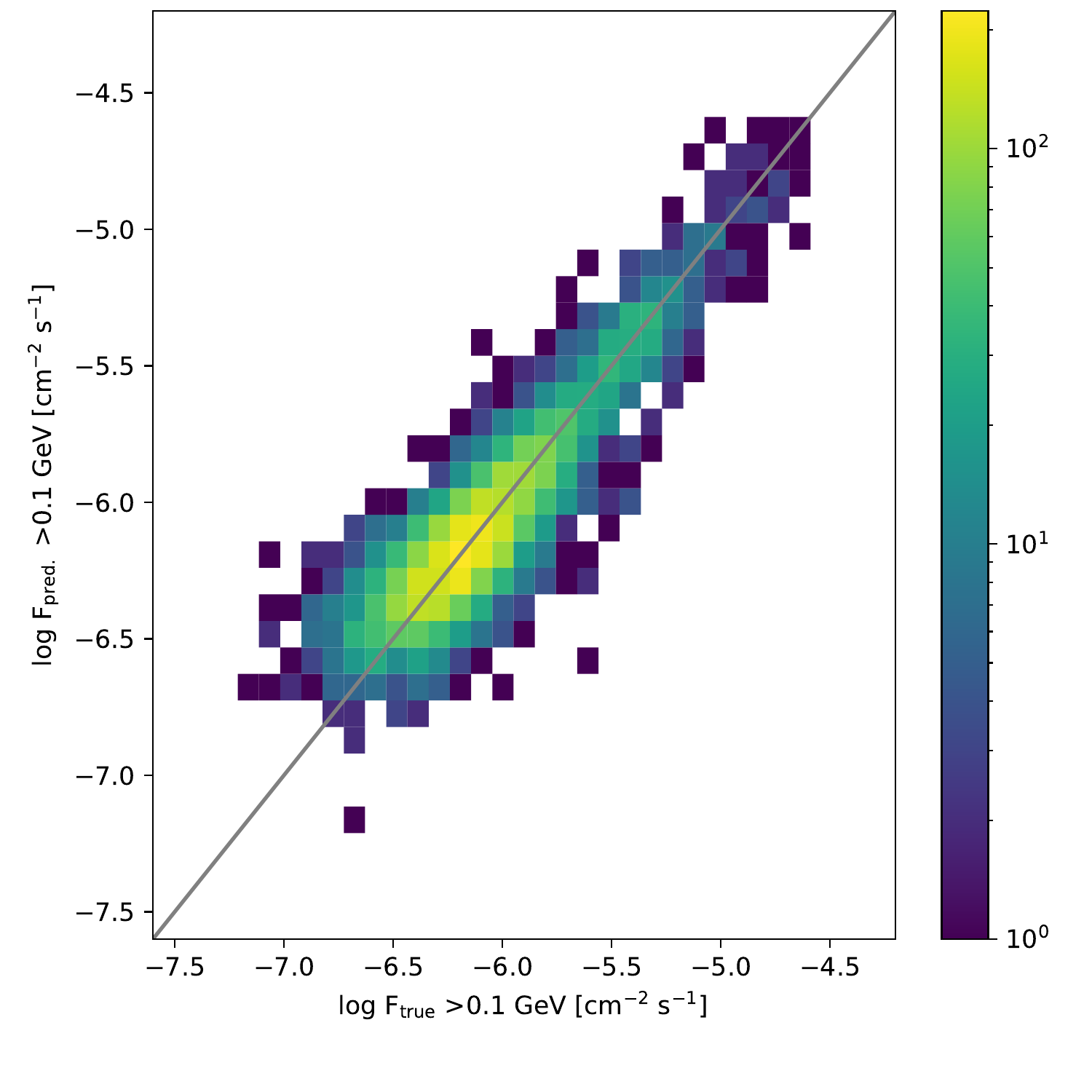}
\includegraphics[width=0.33\textwidth]{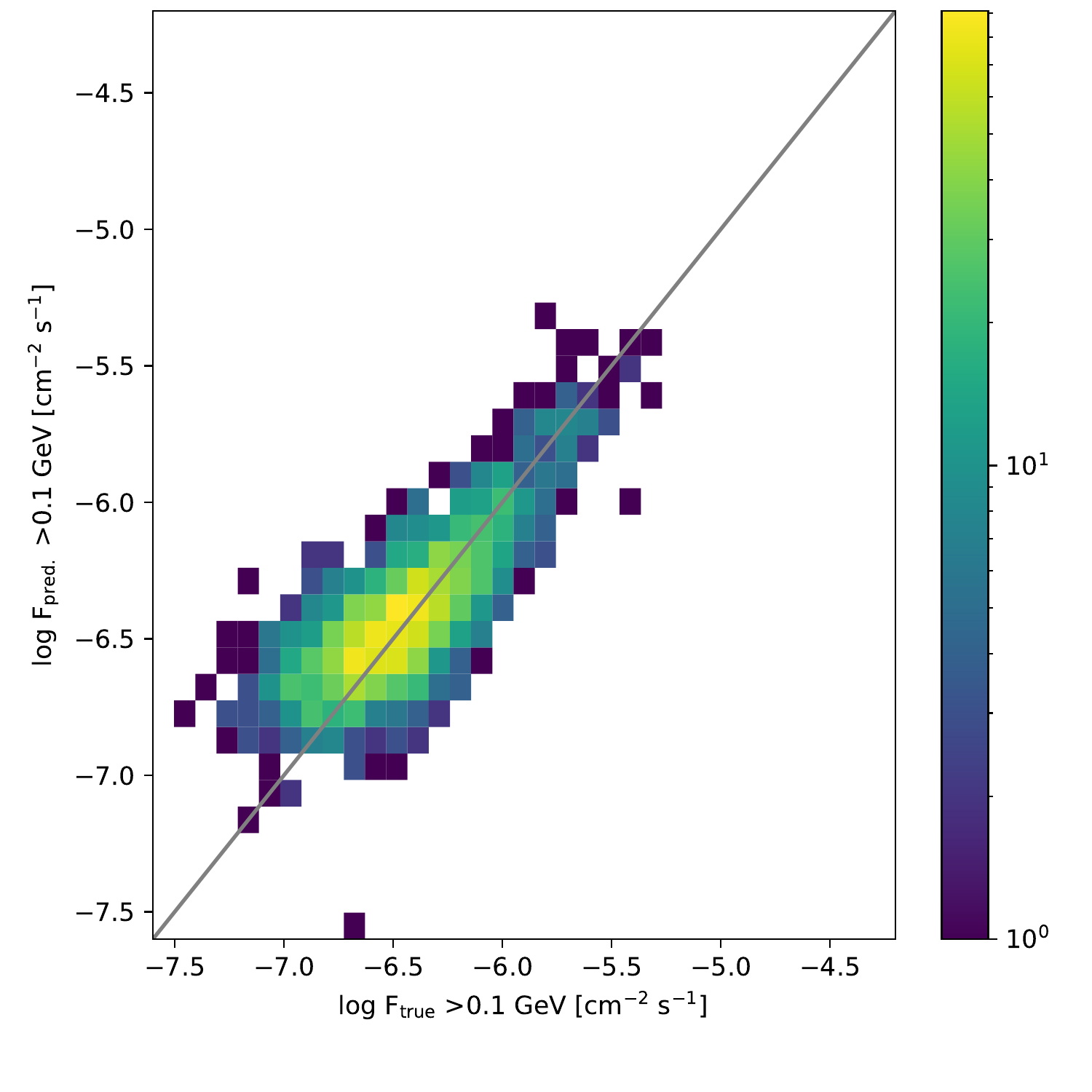}
\includegraphics[width=0.33\textwidth]{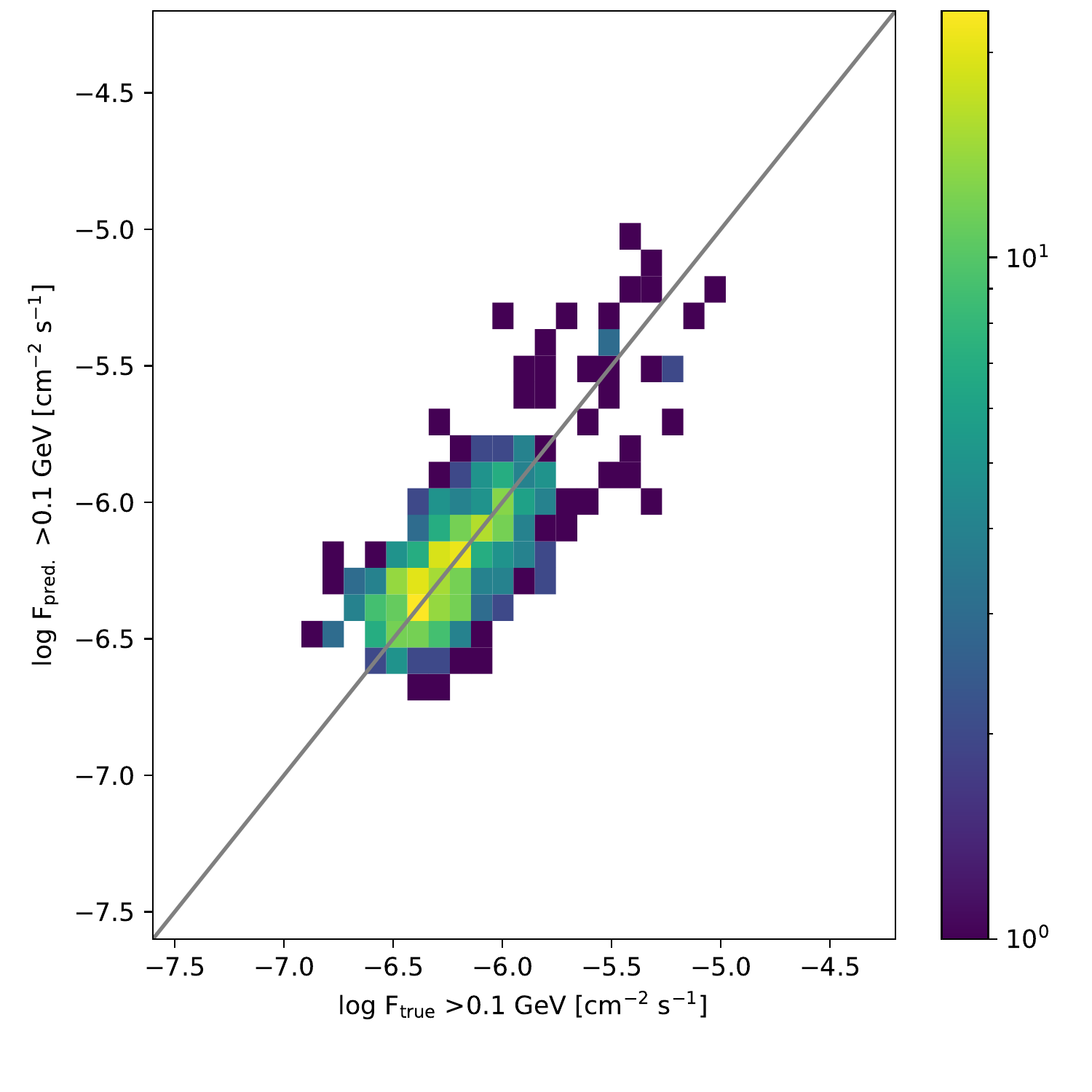}
  \caption{
  Distribution of predicted vs measured value of flux for direct training (top panel) and training on differences (bottom panel) for different source classes: FSRQ (left), BLL (middle) and other (right). 
  The points for which either the last available measurement or the measurement that is being predicted resulted in an upper limit are omitted.}\label{fig5}
\end{figure*}

We also divide the sources into types following Table~\ref{tab:srcs}. 
For each type we combine all sources of a given type into one sample (as in (B) training)  and perform training on such a joint sample. 
The comparison of the predicted and true flux values is shown in Fig.~\ref{fig5}, while the  performance measures are summarised in Table~\ref{tab:bytype}.
\begin{table*}[t!]
  \centering
  \begin{tabular}{l|l|r|r|r|r | r|r|r|r }
   \multicolumn{2}{c}{} & \multicolumn{4}{|c|}{TS} & \multicolumn{4}{|c}{$F$}  \\ \hline
    Train method & Train sample & MAE   & $f_{u/d}$[\%] & $f_u$[\%] & $f_d$[\%] & MAE & $f_{u/d}$[\%] & $f_u$[\%] & $f_d$[\%] \\\hline
%    (i) Direct       & FSRQ & 0.732 & $74.1\pm0.2$ & $66.1\pm0.3$ & $82.1\pm0.3$ & 0.752 & $73.0\pm0.5$ & $48.4\pm0.7$ & $97.8\pm0.2$ \\
%    (i) Direct       & BLL  & 0.877 & $73.1\pm0.3$ & $63.7\pm0.5$ & $82.7\pm0.4$ & 0.936 & $80.4\pm0.5$ & $62.0\pm0.9$ & $98.9\pm0.2$ \\
%    (i) Direct       & other& 0.688 & $75.6\pm0.6$ & $68.3\pm0.9$ & $82.9\pm0.8$ & 0.990 & $77.8\pm1.3$ & $56.1\pm2.2$ & $99.4\pm0.3$ \\\hline
%    (ii) Differences & FSRQ & 0.836 & $71.9\pm0.2$ & $79.5\pm0.3$ & $64.4\pm0.3$ & 0.336 & $72.2\pm0.5$ & $63.8\pm0.7$ & $80.7\pm0.6$ \\
%    (ii) Differences & BLL  & 0.940 & $70.9\pm0.3$ & $79.5\pm0.4$ & $62.3\pm0.5$ & 0.447 & $77.2\pm0.6$ & $74.0\pm0.8$ & $80.4\pm0.7$ \\
%    (ii) Differences & other& 0.834 & $72.4\pm0.6$ & $80.4\pm0.8$ & $64.5\pm1.0$ & 0.393 & $74.6\pm1.4$ & $66.5\pm2.1$ & $82.8\pm1.7$ \\
    (i) Direct       & FSRQ & 0.732 & $74.1\pm0.2$ & $66.1\pm0.3$ & $82.1\pm0.3$ & 0.289 & $63.6\pm0.7$ & $52.3\pm1.0$ & $75.1\pm0.8$ \\
    (i) Direct       & BLL  & 0.877 & $73.1\pm0.3$ & $63.7\pm0.5$ & $82.7\pm0.4$ & 0.353 & $69.1\pm1.0$ & $65.5\pm1.4$ & $72.8\pm1.3$ \\
    (i) Direct       & other& 0.688 & $75.6\pm0.6$ & $68.3\pm0.9$ & $82.9\pm0.8$ & 0.346 & $62.6\pm2.3$ & $52.2\pm3.3$ & $72.9\pm2.9$ \\\hline
    (ii) Differences & FSRQ & 0.836 & $71.9\pm0.2$ & $79.5\pm0.3$ & $64.4\pm0.3$ & 0.279 & $63.2\pm0.7$ & $70.0\pm0.9$ & $56.3\pm1.0$ \\
    (ii) Differences & BLL  & 0.940 & $70.9\pm0.3$ & $79.5\pm0.4$ & $62.3\pm0.5$ & 0.344 & $65.3\pm1.0$ & $81.1\pm1.2$ & $49.5\pm1.5$ \\
    (ii) Differences & other& 0.834 & $72.4\pm0.6$ & $80.4\pm0.8$ & $64.5\pm1.0$ & 0.326 & $60.7\pm2.3$ & $70.4\pm3.0$ & $51.1\pm3.3$ \\
  \end{tabular}
  \caption{Comparison of the performance parameters for different types of sources.}
  \label{tab:bytype}    
  \end{table*}
In general the predictions for the individual source classes follow the properties observed in training of the full sample. 
The differences between those classes can stem from difference of observed brightness (e.g. FSRQ sources are responsible for the highest registered states in the sample), and from intrinsic properties (variability) of the source class. 
The accuracy of flux prediction measured with MAE parameter is best for FSRQ. 
On the other hand BL Lac sources have the highest fraction of properly predicted flux light curve trend, $f_{u/d}$. 
%The lowest value of $f_{u/d}$ for flux interestingly is achieved for FSRQ.
%Comparing FSRQ with BL Lac objects: better flux prediction combined with the worse trend prediction for the former class is in line with the fact that on one hand those sources are in general bright, however at the same time violently variable. 
Better flux prediction combined with the worse trend prediction for the FSRQ class than with BL Lac objects is in line with the fact that on one hand those sources are in general bright, however at the same time violently variable (on sub-day time scales). 
The ''other'' class of sources is in general not very bright and also poorly populated with significant flux measurements. This likely causes worse MAE and $f_{u/d}$ accuracy in this class.

%\afterpage{\clearpage}

\subsection{Prediction over longer time scale}
\begin{figure}[t!]
    \includegraphics[width=0.48\textwidth]{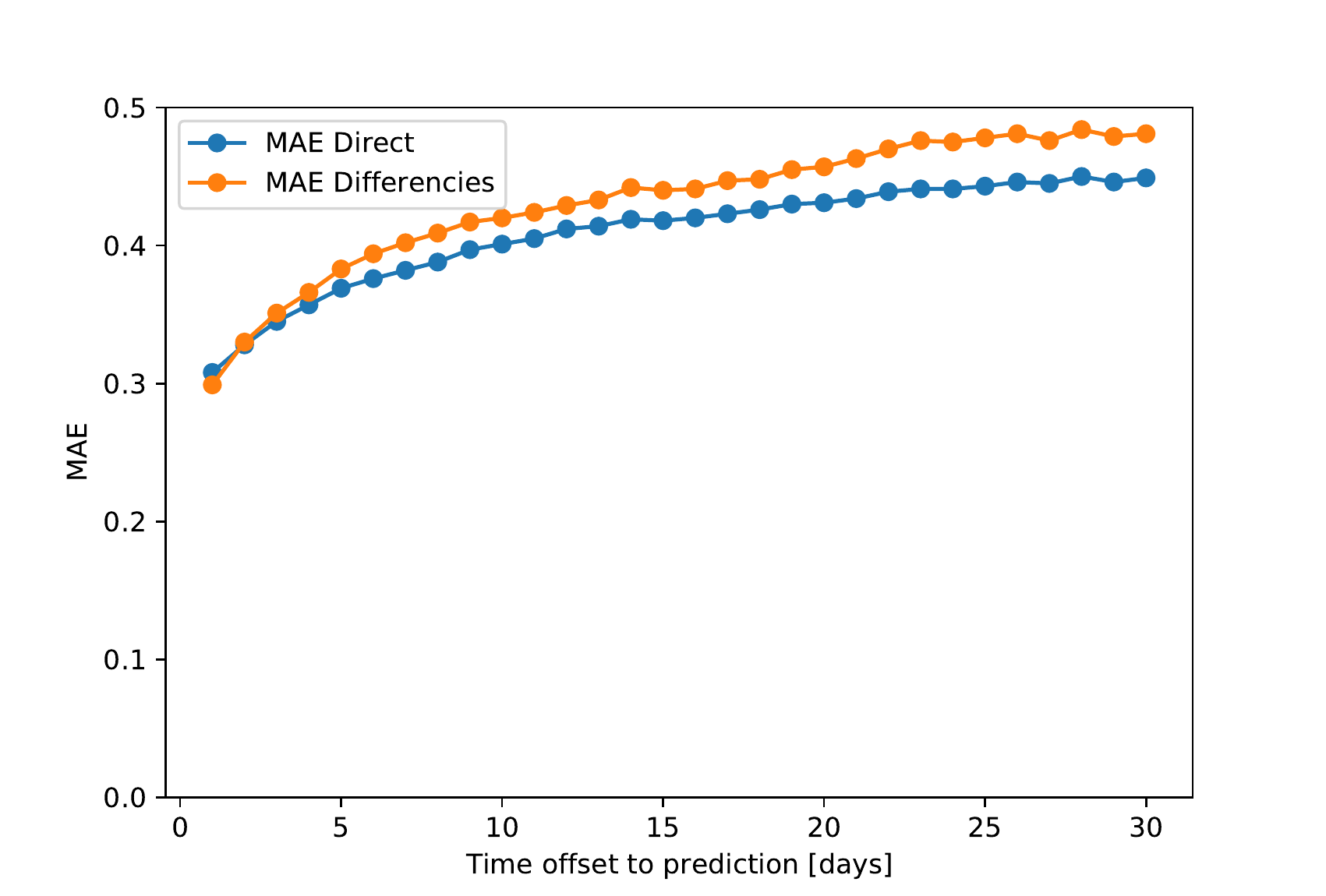}
  \caption{Dependence of flux accuracy on the number of days to predict in future. Training over the full sample of sources. }\label{fig7}
\end{figure}

\begin{figure}[t!]
    \includegraphics[width=0.48\textwidth]{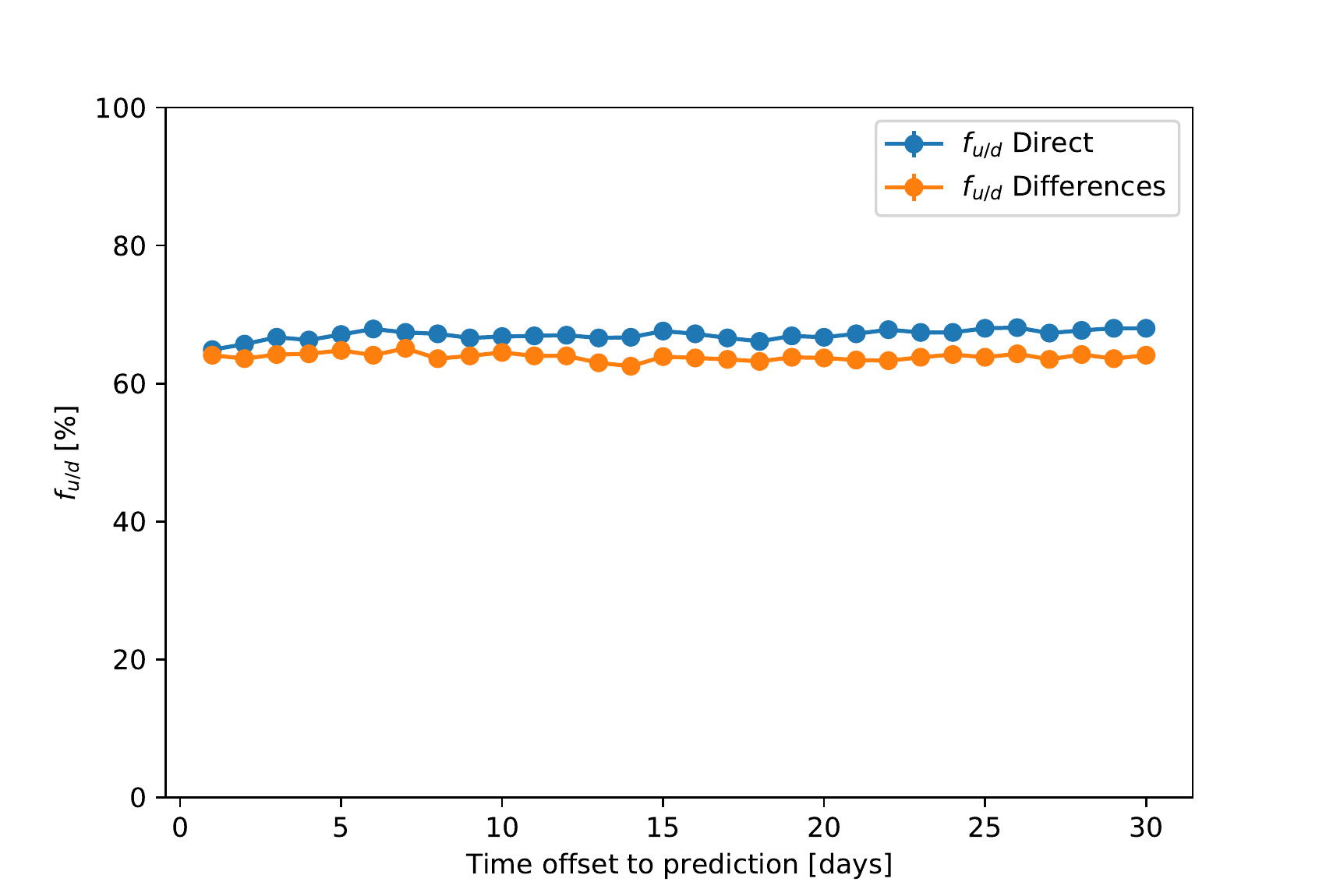}
  \caption{Dependence of flux trend prediction accuracy on the number of days to predict in future. Training over the full sample of sources. }\label{fig8}
\end{figure}

In order to further test the validity of the method %and to understand the underlying prediction principle 
we tested how the performance measures of flux prediction vary if the time offset to next measurement is increased. 
In Fig.~\ref{fig7} and Fig.~\ref{fig8} we show how MAE and $f_{u/d}$ of flux  evolve if the prediction is done over a number of days in future.
The training is done over the whole sample of sources together ((B) training scheme). 
We use the same quality cuts in the validation, namely take into account only the points for which both the actual measurement which is being predicted and the last available measurements used in the prediction are not upper limits. 

In the case of MAE, as expected, the prediction power degrades with the raising time offset to the last available measurement. 
The effect is most strongly pronounced over the first few days, and somewhat more pronounced in the case of ''differences'' training. 
This is in line with the results of Fig.~\ref{fig1} that showed that mainly the last few measurements are used in the prediction. 

The dependence of $f_{u/d}$ is more surprising. The fractions of properly predicted raises and drops of flux is nearly independent on the number of days over which the prediction is done. 
This shows that the prediction of RF on the source trend is mainly dependent on the global properties of the tested sources (such as the characteristic time scales of flux increase or decrease or typical flux of the source), rather than on the most recent behaviour of the source. 
The derived values of MAE for flux prediction result in accuracy of a few tens of per cent. 
Such accuracy will often not be sufficient when compared to a point-to-point measured flux changes, especially that they are additionally burdened by instrumental uncertainties of the measured fluxes. 
This further explains why the prediction of the flux increase/decrease would be more connected with average behaviour of the sources (which constitutes the training sample) and thus not depend strongly on the time offset to the last measurement. 
For ''direct'' training, there is a small improvement of the $f_{u/d}$ between the prediction to the next day and prediction over a few days to the future. 
This we attribute to a possible sample selection bias. 
Namely, as the cases in which either the last available measurement or the actual predicted measurement are upper limits are removed from the evaluation, such a condition for a longer time offsets will more efficiently remove a higher fraction of short duration flares that are likely more stochastic in nature and thus intrinsically more difficult to predict.

\section{Discussion and conclusions} \label{sec:conc}
Using machine learning techniques we tested if it is possible to predict the state of an AGN object and the trend of its emission over a short time scale.
We used a RF method to predict flux and TS evolution of GeV emission observed from a selected sample of 88 AGN objects monitored by \fermilat{}. 
Two training schemes are tested, ''direct'' testing on the measured values, and ''differences'' training on the daily changes of the training variable. 
%We show that a partial prediction is possible. 
If the prediction is done over a single day time scale, then the proposed method achieved an accuracy of about a factor of two for TS values above 1.
Flux above 0.1 GeV can be predicted over the same time scale with an accuracy of 35\% measured over the points that resulted in measurable flux, as long as the last available measurement is also not an upper limit. 
Separating the sample according to the source class does not show big differences between different types of objects, however the prediction of the value is slightly more accurate for FSRQ, which are also the brightest and the most populated class in the sample. 
As expected the accuracy of the prediction degrades over longer time scales, however not very strongly.  
We show that a partial prediction (with $\sim60-75\%$ accuracy) of the raising/falling emission trends is possible. 
Surprisingly, a similar fraction is achieved also if prediction is done over a larger number of days into future, strongly suggesting that the underlying basis of the prediction is not in the recent emission of the source, but rather in the general characteristic of those sources (e.g. typical raising/fall time scales, typical fluxes). 
%The obtained prediction power is also dependent on the sample selection, in particular on the treatment of the flux upper limits. 

Likely due to limited size of samples, training over combined samples from many sources turned out to be superior to individual training on each source separately.  
Both training schemes ''direct'' and ''differences'' resulted in comparable performance. 
However, while the total fraction of properly predicted trends is similar, ''direct'' training is more efficient at predicting raises, while ''differencies'' on drops. 
Testing different training schemes (inclusion of flux uncertainties in the training, usage of a different machine learning method) did not improve the results considerably, suggesting that the prediction power is limited by the source properties (intrinsic randomness), and/or instrumental limitations (daily sensitivity of \fermilat{}). 

The proposed method has some limitations. 
In addition to the above-mentioned biases, the presence of upper limits in the sample complicates the training. 
Also the prediction power is likely affected by the limited statistical accuracy of one-day measurements with \fermilat{}.
Nevertheless, it is interesting that the combination of deterministic and random processes governing the emission of AGN objects can be used for actual prediction of the flux measurements. 
This is in line with Ornstein-Uhlenbeck-like processes governing the variability of AGN objects \citep{bu21}. 
The already gathered nearly 13 years of data with the \fermilat{} instrument provide a sufficient sample for training on emission trends, in particular if statistics of various sources are combined together. 
The proposed method might provide an additional information in automatic selection of targets for ToO follow-up with Cherenkov telescopes, in particular the upcoming CTA.

\section*{Acknowledgments}
This work is supported by the grant through the Polish National Research Centre No. 2019/34/E/ST9/00224.
We would like to thank D. Sobczy\'nska and two anonymous journal reviewers for their comments that helped to improve the manuscript. 





\appendix

\section{Optimisation of the machine learning method}\label{sec:opt}
Using the full sample of sources and training on fluxes we performed further tests to see if the performance of the prediction can be improved.
The performance measures for different tests are summarised in Table~\ref{tab:pars_tests}.
\begin{table*}[t]
  \centering
  \begin{tabular}{l|l | r|r|r|r }
%    \multicolumn{2}{c|}{Train method}  
Train method & Enhancement
    & MAE & $f_{u/d}$[\%] & $f_u$[\%] & $f_d$[\%] \\\hline
    (i) Direct   &  \textbf{reference} &  0.308 & $64.9\pm0.5$ & $56.5\pm0.8$ & $73.4\pm0.7$ \\
    (i) Direct   &  no train U.L. &  0.298 & $65.2\pm0.5$ & $63.9\pm0.8$ & $66.5\pm0.8$ \\
    (i) Direct   &  with uncertainties &  0.302 & $65.3\pm0.5$ & $58.2\pm0.8$ & $72.5\pm0.7$ \\
    (i) Direct   &  unc.+no U.L. train &  0.298 & $65.4\pm0.5$ & $62.7\pm0.8$ & $68.2\pm0.7$ \\\hline
    (i) Direct   &  MLPRegressor &  0.308 &  $65.7\pm0.5$ & $64.4\pm0.8$ & $66.9\pm0.7$\\\hline\hline
    (ii) Differences   &  \textbf{reference}  & 0.299 & $64.1\pm0.5$ & $72.6\pm0.7$ & $55.5\pm0.8$ \\
    (ii) Differences   &  no train U.L.   & 0.304 &  $63.2\pm0.5$ & $71.3\pm0.7$ & $55.0\pm0.8$ \\
    (ii) Differences   & with uncertainties & 0.310 & $61.7\pm0.5$ & $79.3\pm0.6$ & $44.1\pm0.8$ \\
    (ii) Differences   & unc.+no U.L. train & 0.307 & $62.5\pm0.5$ & $72.7\pm0.7$ & $52.3\pm0.8$ \\\hline
    (ii) Differences   & MLPRegressor & 0.299 & $64.6\pm0.5$ & $70.5\pm0.7$ & $58.7\pm0.8$ 
  \end{tabular}
  \caption{Comparison of the performance parameters obtained for training enhancements. Training is done on the flux variable using the complete data set.}
  \label{tab:pars_tests}    
  \end{table*}

\subsection{Bias of upper limits in the training}\label{sec:noul}
For a fair evaluation of prediction power we are limiting the calculation of performance measures to only those where neither the predicted measurement nor last available measurements are an upper limit. 
Such a condition is however not applied to the training sample, which on one hand increases the available statistics for training, but can as well produce biases in the training. 
To evaluate the effect we perform a test excluding such events from the training as well. 
The performance measures in Table~\ref{tab:pars_tests} show that the MAE and $f_{u/d}$ values are not changed significantly. 
There is an interesting effect on the fraction of the properly predicted trends in the case of ''direct'' training: it improves for raises and worsens for decreases.  
Interestingly both $f_u$ and $f_d$ are becoming similar for ''direct'' training, however they remain considerably different for ''differences'' training. 

\subsection{Usage of flux uncertainties}\label{sec:uncert}

\begin{figure}[t]
  \includegraphics[width=0.48\textwidth]{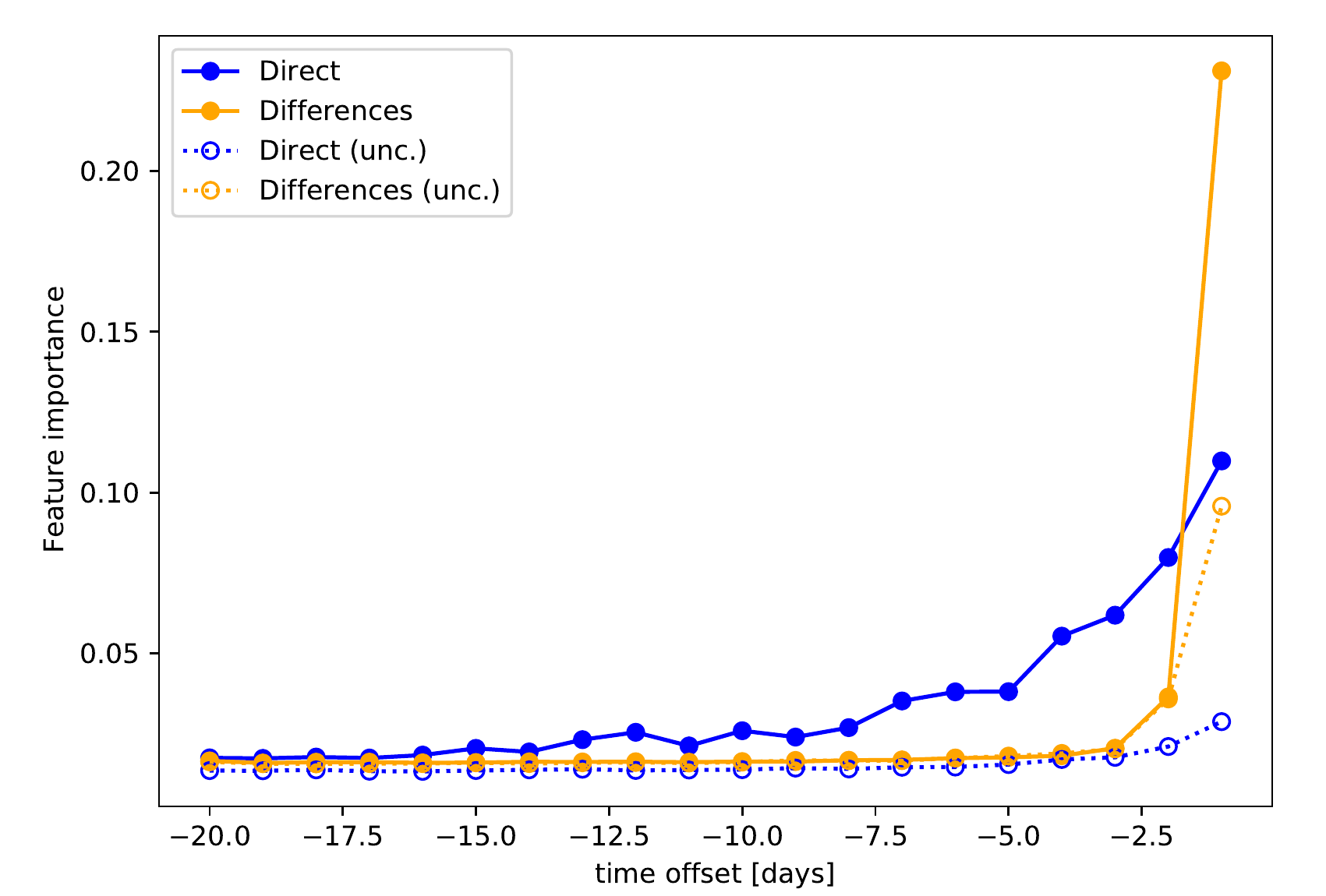}
  \caption{Relative importance of individual parameters used in the training including also uncertainties of the measurements (the values are shown with solid markers and lines, while uncertainties of the values are shown with empty markers and dotted lines).
  Training is done for flux values in the full sample of AGN objects on direct values (blue) or differences (orange). 
  }\label{fig:rf_unc}
\end{figure}

In the previous sections only the flux values were used in the training, neglecting their uncertainties.
The flux uncertainties of the previous measurements bring additional information that can be exploited in the training. 
In order to test this we introduce a different training scheme. 
In the ''direct'' training, in addition to the used $H$ earlier measurements of the flux as the input variables of the RF, we add another $H$ variables corresponding to the statistical uncertainties of these fluxes. 
For ''differences'' type of training we propagate the uncertainty of the individual fluxes to estimate the uncertainty of individual flux differences. 
If a given flux measurement is an upper limit, its uncertainty is set to the upper limit value. 
To normalise the range of the input variables we use the same transformation as in Eq.~\ref{eq:trans1} also for those uncertainties. 
The relative importance of all the input parameters of the RF are shown in Fig.~\ref{fig:rf_unc}.
Similarly to the case of the flux values, the most recent flux uncertainties have the highest importance in the training.
Nevertheless the relative importance of the uncertainties is smaller (in particular for the case of ''direct'' training) than of the fluxes. 
The lesser effect of the uncertainties in the training can be understood as the relative accuracy of the uncertainties is usually worse than the relative accuracy of the measurement. 
Moreover, since in the used data set the integration times of each measurement are the same, and hence the performance of the instrument for each observation is similar, the obtained flux values and uncertainties are correlated. 
In Table~\ref{tab:pars_tests} we compare the performance measures for the training of flux neglecting or using the uncertainties. 

The achieved MAE values are within 4\% of the ones obtained in the reference training. 
For the ''direct'' training the fractions of the properly predicted raises and drops of the flux is not strongly affected by inclusion of uncertainties in the training scheme. 
For the ''differences'' training however there is a large increase of properly predicted raises and decrease of properly decreased drops. 
It is likely connected by a bias connected with presence of flux upper limits in the training sample and causes a net drop of the properly predicted behaviour by $\sim2.4\%$.
Consistently with this interpretation if additionally removal of upper limit following \ref{sec:noul} is combined with usage of uncertainties the $f_u$ and $f_d$ values are closer to the case of not using flux uncertainties in the training. 

\section{Neural networks}\label{sec:nn}
\begin{figure*}[t!]
\includegraphics[width=0.48\textwidth]{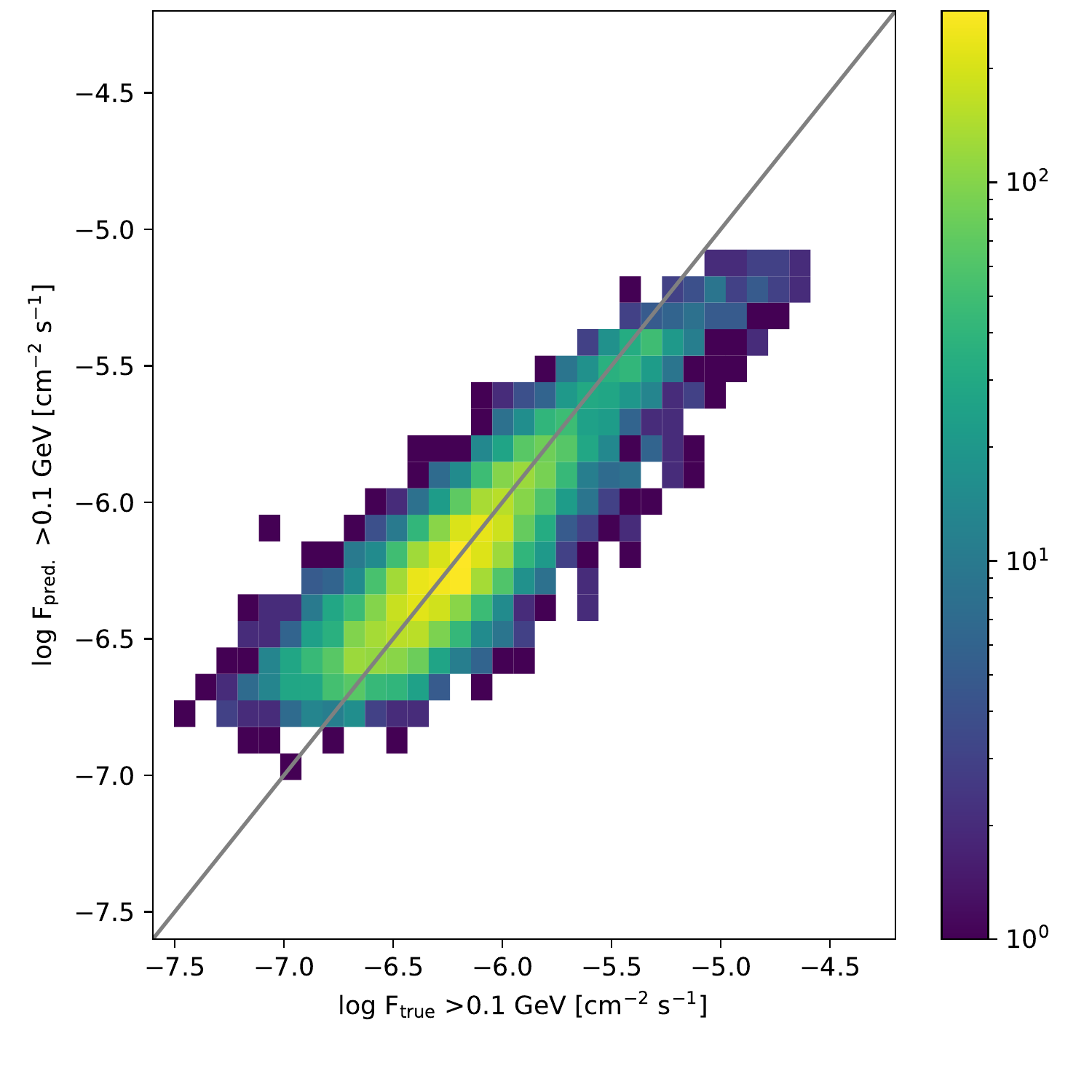}
\includegraphics[width=0.48\textwidth]{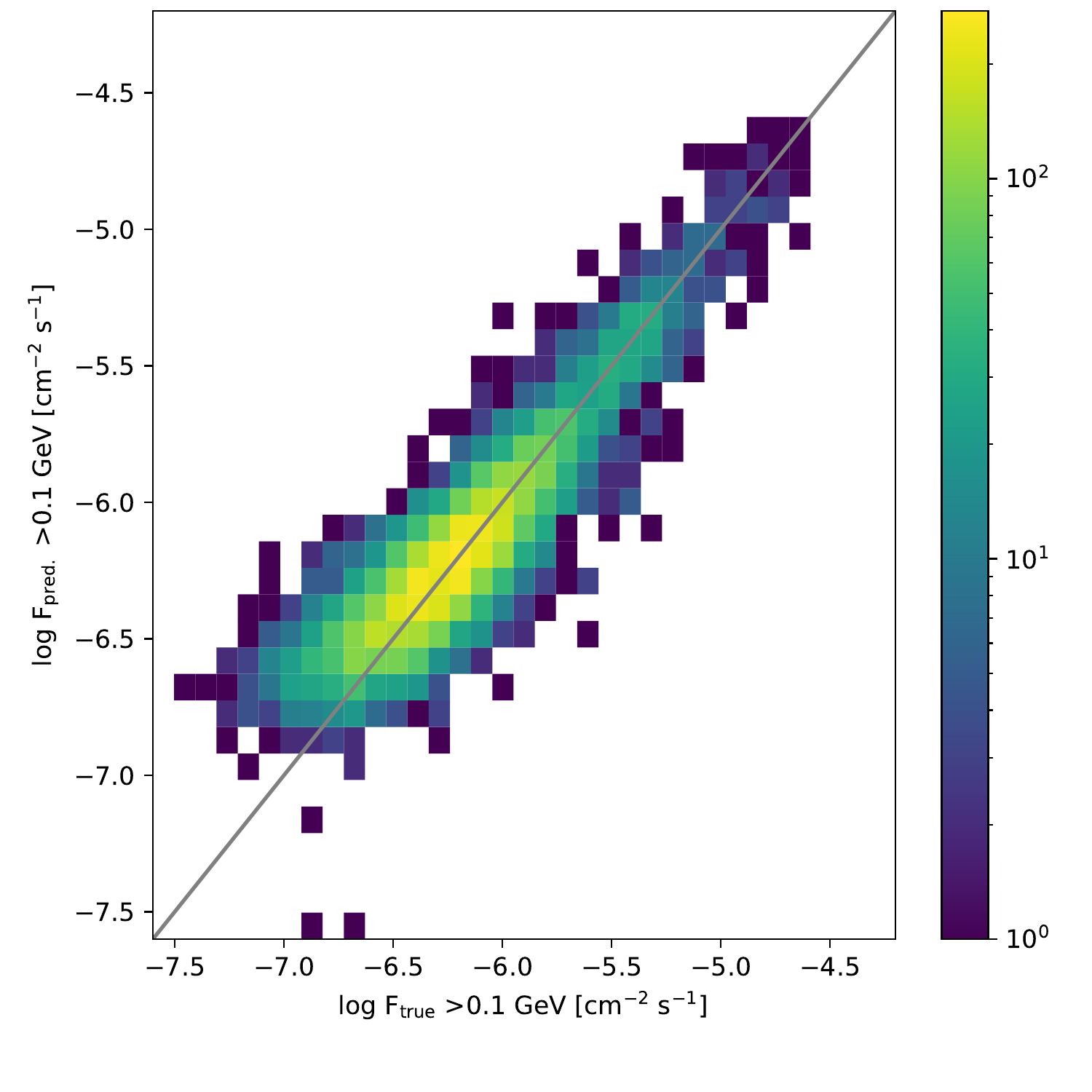}
  \caption{
  Distribution of predicted vs measured value of flux for MLPRegressor direct training (left panel) and training on differences (right panel) for the whole sample. 
  The points for which either the last available measurement or the measurement that is being predicted resulted in an upper limit are omitted.}\label{fig:nn}
\end{figure*}
In order to test if the prediction power is limited by the machine learning method used, rather than by intrinsic randomness and limitations of the training sample we test the usage of neural networks.
We use  Multi-layer Perceptron regressor (MLPRegressor) \citep{hi89} with ''adam'' minimisation \citep{2014arXiv1412.6980K} implemented in the \texttt{scikit-learn} package \citep{pe11}.  
While Eqs.~\ref{eq:trans1} and ~\ref{eq:trans2} already can be considered to be the activation function of the neural network, we saw a small improvements in the performance measures if $\tanh$ is additionally used.
We use one hidden layer with 100 neurons. 
We have tested also a single hidden layer with 30 or 300 neurons and two hidden layers with 10 and 5 neurons, all of them gave comparable results to the baseline model with 100 neurons. 
The training is stopped when the loss function is not improving by more then $10^{-4}$ in the last 10 iterations. 
For the tested case of training on the full sample this is achieved after 76 iterations for ''direct'' training and 176 for ''differences'' training. 
The training score is defined as $R^2=(1-u/v)$, where $u$ is the residual sum of squares and $v$ is the variance of the test sample true values ($R^2 = 1$ for perfect prediction and $R^2=0$ for a  constant model that always predicts an average value of the sample, $R^2$ can be also arbitrarily negative).
The ''direct'' and ''differences'' training reached $R^2$ value of 0.454 and 0.244 respectively. 
The predicted values are confronted with the measured ones in Fig.~\ref{fig:nn}, and the performance measures are summarised in Table~\ref{tab:pars_tests}.
The distribution of the predicted vs measured values shows similar features to the one obtained for the RF. 
Also the performance measures are similar.
This suggests that the prediction power is not limited by the machine learning method, but rather by the intrinsic ''randomness'' of the studied sources, \fermilat{} performance and possibly size of the training samples.

\end{document}